\documentclass[useAMS,usenatbib,usegraphicx]{mn2e}
\usepackage{graphicx,times,epsfig,rotating}

\def\mnras{MNRAS}
\def\aj{AJ}
\def\aap{A\&A}
\def\apj{ApJ}
\def\apjl{ApJ}
\def\apjs{ApJS}
\def\araa{ARA\&A}

\def\nat{Nature}

\newcommand{\msun}{\mbox{M$_{\sun}$ }}
\newcommand{\msunend}{\mbox{M$_{\sun}$}}
\newcommand{\lsun}{\mbox{L$_{\sun}$ }}
\newcommand{\lsunend}{\mbox{L$_{\sun}$}}
\newcommand{\lir}{\mbox{L$_{\rm IR}$}}

\newcommand{\cmthree}{\mbox{cm$^{-3}$}}

\newcommand{\msunyr}{\mbox{M$_{\sun}$yr$^{-1}$ }}
\newcommand{\msunyrend}{\mbox{M$_{\sun}$yr$^{-1}$}}
\newcommand{\htwo}{\mbox{H$_2$}}

\newcommand{\z}{\mbox{$z$}}
\newcommand{\zapprox}{\mbox{$z \approx $}}
\newcommand{\zga}{\mbox{$z \ga$}}

\newcommand{\zsim}{\mbox{$z\sim$ }}

\newcommand{\lbol}{\mbox{L$_{\rm bol}$}}

\newcommand{\sef}{\mbox{$F_{\rm 850}$}}

\newcommand{\sunrise}{\mbox{\sc sunrise}}
\newcommand{\gadget}{\mbox{\sc gadget-3}}
\newcommand{\starburst}{\mbox{\sc starburst99}}

\newcommand{\bzk}{\mbox{{\it BzK}}}

\newcommand{\xco}{\mbox{$X_{\rm CO}$}}

\newcommand{\mc}{\mbox{$\hat{M}_{\rm c}$}}
\newcommand{\msunpcsq}{\mbox{$\msun {\rm pc}^{-2}$}}
\newcommand{\mjchar}{\mbox{$\hat{M_{\rm J}}$}}
\newcommand{\fsps}{\mbox{\sc fsps}}
\newcommand{\arepo}{\mbox{\sc arepo}}

\title[Implications of an IMF that Varies with the Jeans Mass in Galaxies]{Cosmological implications of a stellar initial mass function that varies with the Jeans mass in galaxies}

\author[D. Narayanan \& R. Dav\'e] {Desika\,
  Narayanan$^{1}$\thanks{E-mail:
    dnarayanan@as.arizona.edu}\thanks{Bart J Bok Fellow} \& Romeel
  Dav\'e$^{1}$\\ $^{1}$Steward Observatory,
  University of Arizona, 933 N Cherry Ave, Tucson, Az,
  85721}
\begin{document}

\date{MNRAS Accepted 2012 April 21st.  Received 2012 April 19th; in
  original form 2012 February 29th. }

\pubyear{2010}

\maketitle

\label{firstpage}

\begin{abstract}

Observations of star-forming galaxies at high-\z\ have suggested
discrepancies in the inferred star formation rates (SFRs) either
between data and models, or between complementary measures of the SFR.
These putative discrepancies could all be alleviated if the stellar
initial mass function (IMF) is systematically weighted toward more
high-mass star formation in rapidly star-forming galaxies.  Here, we
explore how the IMF might vary under the central assumption that the
turnover mass in the IMF, \mc,\ scales with the Jeans mass in giant
molecular clouds (GMCs), \mjchar.  We employ hydrodynamic simulations
of galaxies coupled with radiative transfer models to predict how the
typical GMC Jeans mass, and hence the IMF, varies with galaxy
properties.  We then study the impact of such an IMF on the star
formation law, the SFR-$M_*$ relation, sub-millimetre galaxies (SMGs),
and the cosmic star formation rate density.  Our main results are: The
\htwo\ mass-weighted Jeans mass in a galaxy scales well with the SFR
when the SFR is greater a few $M_\odot$~yr$^{-1}$.  Stellar population
synthesis modeling shows that this results in a nonlinear relation
between SFR and \lbol, such that SFR $\propto \lbol^{0.88}$.  Using
this model relation, the inferred SFR of local ultraluminous infrared
galaxies decreases by a factor $\sim\times 2$, and that of
high-\z\ SMGs decreases by $\sim\times 3-5$.  At \zsim 2, this results
in a lowered normalisation of the SFR-$M_*$ relation in better
agreement with models, a reduced discrepancy between the observed
cosmic SFR density and stellar mass density evolution, and SMG SFRs
that are easier to accommodate in current hierarchical structure
formation models.  It further results in a Kennicutt-Schmidt (KS) star
formation law with slope of $\sim 1.6$ when utilising a physically
motivated form for the CO-\htwo\ conversion factor that varies with
galaxy physical property.  While each of the discrepancies considered
here could be alleviated without appealing to a varying IMF, the
modest variation implied by assuming \mc$\propto$\mjchar\ is a
plausible solution that simultaneously addresses numerous thorny
issues regarding the SFRs of high-\z \ galaxies.

\end{abstract}
\begin{keywords}
stars:luminosity function, mass function -- stars: formation --
galaxies: formation --galaxies: high-redshift -- galaxies: ISM --
galaxies: starburst -- cosmology:theory
\end{keywords}

\section{Introduction}
\label{section:introduction}

The buildup of stellar mass over cosmic time is a central issue in
understanding the formation and evolution of galaxies.  A common approach
to quantifying stellar growth is to measure the evolution of the star
formation rates (SFRs) of galaxies.  This is done using a wide variety
of tracers from the ultraviolet (UV) to the radio.  Generally, all such
measures trace the formation rate of higher-mass (typically O and B)
stars, while the bulk of the stellar mass forming in lower-mass stars
is not directly detected.  Hence measuring the true rate of stellar
growth requires assuming a conversion between the particular tracer
flux and the total stellar mass being generated~\citep[e.g.][]{ken98a,ken12}.
This requires assuming some stellar initial mass function (IMF), namely
the number of stars being formed as a function of mass.

On global cosmological scales, multi-wavelength observations of
galaxies are converging on a broad scenario for the cosmic star
formation rate evolution~\citep{mad96}.  Galaxies at high-redshift
appear to be more gas-rich and forming stars more rapidly at a given
stellar mass than present day galaxies \citep[see the recent review
by][]{sha11}.  The cosmic star formation rate density rises slowly
from early epochs to peak between redshifts \zapprox 1-3, and then
declines toward \z=0 \citep[e.g. ][]{hop06d}.  This global evolution
is also reflected in measurements of the star formation rates of
individual galaxies, which grow significantly at a given stellar
mass from today out to $z\sim 2$, prior to which they show a much
slower evolution~\citep[e.g.][]{dad05,dav08,gon10,hop10}.

Meanwhile, recent advances in near-infrared capabilities have enabled
measurements of the buildup of stellar mass out to high redshifts,
as traced by more long-lived stars (typically red giants).  Nominally,
the integral of the cosmic star formation rate, when corrected for
stellar evolution processes, should yield the present-day cosmic
stellar mass.  Analogously, the time differential of the stellar
mass evolution should be the same as measured star formation rate.
Thus in principle there is now a cross-check on the rate of high-mass
stars forming relative to lower-mass stars.

Preliminary comparisons along these lines have yielded general
agreement out to $z\sim 1$~\citep[e.g.][]{bel07b}.  However, moving to
higher redshifts into the peak epoch of cosmic star formation, there
are growing hints of discrepancies: the integrated cosmic star
formation rate density seems to exceed the observed stellar mass
function~\citep[accounting for stellar mass
  loss;][]{hop06d,els08,wil08c,per08}.  These discrepancies are
relatively mild, at the factor of two to three level, so could perhaps
be resolved by a more careful consideration of systematic
uncertainties in SFR and $M_*$ measures.  Indeed, some analyses fail
to show strong discrepancies~\citep[e.g.][]{sob12}.  Nevertheless, it
is interesting that when discrepancies are seen, they unanimously
favor the idea that the observed stellar mass growth appears slower
than expected from the observed SFR.

Theoretical models have made a number of advances toward understanding
the observed properties of star-forming galaxies at high redshifts.
Simulations advocate a picture in which continual gas accretion
from the intergalactic medium (IGM) feeds galaxies with fresh fuel
\citep[e.g.][]{mo98,rob04,gov09,cer10,age11} and
ultimately drives star formation \citep[][]{ker05,dek08,dek09,ker09}.
In such models, stars typically form at rates proportional to
the global gas accretion rate from the IGM, regulated by feedback
processes~\citep{spr03a,dav11b}.  Such a scenario predicts a fairly tight
and nearly linear relation between star formation rate and stellar mass
for star-forming galaxies~\citep[e.g.][]{dav00,fin06}.  Observations
have now identified and quantified such a relation out to \zsim 6-8
\citep{noe07b,noe07a,dad07,sta09,mcl11}, which has come to be called the
``main sequence'' of galaxy formation.  The bulk of cosmic star formation
appears to occur in galaxies along this main sequence, while merger-driven
starbursts contribute $\la 20\%$ globally~\citep[e.g.][]{rod11,wuy11}.

Although this scenario broadly agrees with observations, a small but
persistent discrepancy has been recently highlighted between the
evolution of the main sequence in simulations and observations,
particularly during the peak epoch of cosmic star formation ($1\la
z\la 3$).  In it, the rate of stellar mass growth at these redshifts
in models is typically smaller by $\sim 2-3\times$ than the observed
star formation rates.  This discrepancy exists for both cosmological
hydrodynamic simulations \citep{dav08}, as well as semi-analytic
models \citep[SAMs;][]{dad07,elb07}, and is to first order independent
of model assumptions about feedback~\citep{dav08}.  This is further
seen in both the global cosmic star formation rate~\citep{wil08c} as
well as when comparing individual galaxies at a given
$M_*$~\citep{dav08}.  
  Similarly, galaxy formation simulations that utilise a variety of
  methods all show a paucity of galaxies that form stars as rapidly as
  galaxies with the highest star formation rates at \zsim 2, the
  sub-millimetre galaxies (SMGs) \citep{bau05,dav10,hay11}.  In all
  cases, the models tend to favor lower true SFRs than implied by
  using available tracers and using conversion factors based on a
  canonical IMF.

One possible but speculative solution to all these discrepancies is
that the stellar IMF in galaxies at $z\sim 2$ is different than what
is measured directly in the Galaxy \citep[e.g. ][]{kan10}.  The
discrepancies described above, between the various observations as
well as between models and data, would all be mitigated by an IMF that
forms somewhat more high-mass stars than low-mass ones at those epochs
compared to the present-day IMF\footnote{This could be described as a
  ``top-heavy" IMF, which we specify as an IMF whose high-mass slope
  is different than local, or a ``bottom-light" IMF, which we define
  as retaining the same high-mass slope but forming fewer low-mass
  stars.  This paper focuses on bottom-light IMFs.}.  Nevertheless, it
is important to point out that at present, no firm evidence that the
IMF varies strongly from the locally observed one \citep[see the
  review by][]{bas10}.  Locally, some observations suggest that a
top-heavy/bottom-light IMF may apply to the Galactic Centre
\citep{nay05,sto05}.  Similarly, \citet{rie93} and \citet{for03}
suggest a turnover mass a factor of $\sim 2-6$ larger than in a
traditional \citet{kro02} IMF in the nearby starburst galaxy M82.
Simultaneous fits to the observed cosmic star formation rate density,
integrated stellar mass measurements, and cosmic background radiation
favour a ``paunchy'' IMF that produces more stars at intermediate
masses \citep{far07}.  \citet{van08} suggested that the IMF may be
more top-heavy at high-redshift (\zapprox 0.8) based on an analysis of
the evolution of the colours and mass-to-light ratios of early-type
galaxies.  However, these observations can all be interpreted without
the need for IMF variations~\citep{bas10}.  Beyond this, some
observations find evidence for a {\it bottom-heavy} IMF in \z=0
early-type galaxies\citep{van11,cap12,con12}.  It is therefore
interesting to examine whether an IMF-based solution is viable and
consistent with a broad suite of observations, both locally and in the
distant Universe.

In this paper, we explore the cosmological consequences of a {\it
  physically-based} model for IMF variations.  Past work has generally
focused on {\it empirically} determining the amount of IMF variation
needed in order to solve one (or more) of the above problems
~\citep[e.g.][]{far07,van08,dav08,wil08c}.  Here, instead, we make a
single critical assumption, first forwarded by Jeans, and later
expanded upon by \citet{lar05} and \citet{tum07}: {\it The IMF
  critical mass (\mc) scales with the Jeans mass in a giant molecular
  cloud} (GMC).  For reference, we call this the Jeans mass
conjecture.  We employ hydrodynamic simulations of isolated galaxies
and mergers including a fully radiative model for the interstellar
medium (ISM) to predict the typical Jeans mass of GMCs in galaxies
with different physical conditions, corresponding to quiescent and
starbursting systems both today and at $z=2$.  Applying the Jeans mass
conjecture, we then make a prediction for how the IMF varies with
global galaxy properties, and explore the implications for such
variations on the discrepancies noted above.  

We emphasise that the main purpose of this paper is to utilise numerical
models of the molecular ISM in galaxies to investigate the consequences
of an IMF in which \mc\ scales with $M_{\rm J}$.  We do not directly
argue for such a scaling relationship; this is taken as an assumption,
and we only seek to study its implications in a cosmological context.
For the reader's edification, we present arguments both for and against
the Jeans mass conjecture in \S~\ref{section:discussion}.

This paper is outlined as follows. In \S~\ref{section:methods}, we
detail our numerical models.  In \S~\ref{section:imf}, we discuss
variations of the IMF versus galaxy star formation rates. In
\S~\ref{section:sfr}, we investigate the effect of this model on
derived star formation rates, paying particular attention to the
star formation law (\S~\ref{section:ks} and
\S~\ref{section:silk}), the SFR-$M_*$ relation at high-redshift
(\S~\ref{section:sfr-mstar}), and the evolution of the cosmic star
formation rate density (\S~\ref{section:madau}).  In
\S~\ref{section:discussion}, we present a discussion, and in
\S~\ref{section:summary}, we summarise.

\section{Numerical Methods}
\label{section:methods}

Our main goal is to simulate the global physical properties of GMCs
in galaxies, and understand the effect of the varying Jeans mass
on observed star formation rates.  These methods and the corresponding
equations are described in significant detail in \citet{nar11b,nar11c};
for the sake of brevity, we summarise the relevant aspects of this
model here, and refer the interested reader to those papers for
further detail.

We first simulate the evolution of galaxies hydrodynamically using
the publicly available code \gadget \ \citep{spr05b,spr05a}.  In
order to investigate a variety of physical environments, we consider
the evolution of disc galaxies in isolation, major (1:1 and 1:3)
mergers, and minor (1:10) mergers at both low (\z=0) and high (\z=2)
redshifts.  These simulations are summarised in Table A1 of
\citet{nar11c}.  The discs are initialised according to the
\citet{mo98} model, and embedded in a live dark matter halo with a
\citet{her90} density profile.  Galaxy mergers are simply mergers
of these discs. The halo concentration and virial radius for a halo
of a given mass is motivated by cosmological $N$-body simulations,
and scaled to match the expected redshift-evolution following
\citet{bul01} and \citet{rob06a}.

For the purposes of the hydrodynamic calculations, the ISM is modeled
as multi-phase, with cold clouds embedded in a hotter phase
\citep{mck77}. The phases exchange mass via radiative cooling of
the hot phase, and supernova heating of cold gas.  Stars form within
the cold gas according to a volumetric Kennicutt-Schmidt star
formation relation, ${\rm SFR} \sim \rho_{\rm cold}^{1.5}$, with
normalisation set to match the locally observed relation
\citep{ken98a,spr00,cox06b}.

In order to model the physical properties of GMCs within these model
galaxies, we perform additional calculations on the SPH models in
post-processing.  We first project the physical properties of the
galaxies onto an adaptive mesh with a $5^3$ base, spanning 200 kpc.
The cells recursively refine in an oct-subdivision based on the
criteria that the relative density variations of metals should be less
than 0.1, and the $V$-band optical depth across the cell $\tau_{\rm V}
< 1$.  The smallest cells in this grid are of order $\sim 70$ pc
across, just resolving massive GMCs.

The GMCs are modeled as spherical and isothermal. The \htwo-HI balance
in these cells is calculated by balancing the photodissociation rates
of \htwo \ against the growth rate on dust grains following the
methodology of \citet{kru08,kru09a} and \citet{kru09b}.  This assumes
equilibrium chemistry for the \htwo.  The GMCs within cells are
assumed to be of constant density, and have a minimum surface density
of 100 \msunpcsq.  This value is motivated by the typical surface
density of local group GMCs \citep{mck07,bol08,fuk10}, and exists to
prevent unphysical conditions in large cells toward the outer regions
of the adaptive mesh.  In practice, the bulk of the GMCs in galaxies
of interest for this study (i.e., galaxies where the galactic
environment has a significant impact on the Jeans mass) have surface
densities above this fiducial threshold value.  With the surface
density of the GMC known, the radius (and consequently mean density)
is known as well. Following \citet{nar11b}, to account for the
turbulent compression of gas, we scale the volumetric densities of the
GMCs by a factor $e^{\sigma_{\rho}^2/2}$, where $\sigma_\rho \approx
{\rm ln}(1+M_{1D}^2/4)$, and $M_{\rm 1D}$ is the 1 dimensional Mach
number of turbulence \citep{ost01,pad02}.

\begin{figure}
\hspace{-1cm}
\includegraphics[scale=0.6]{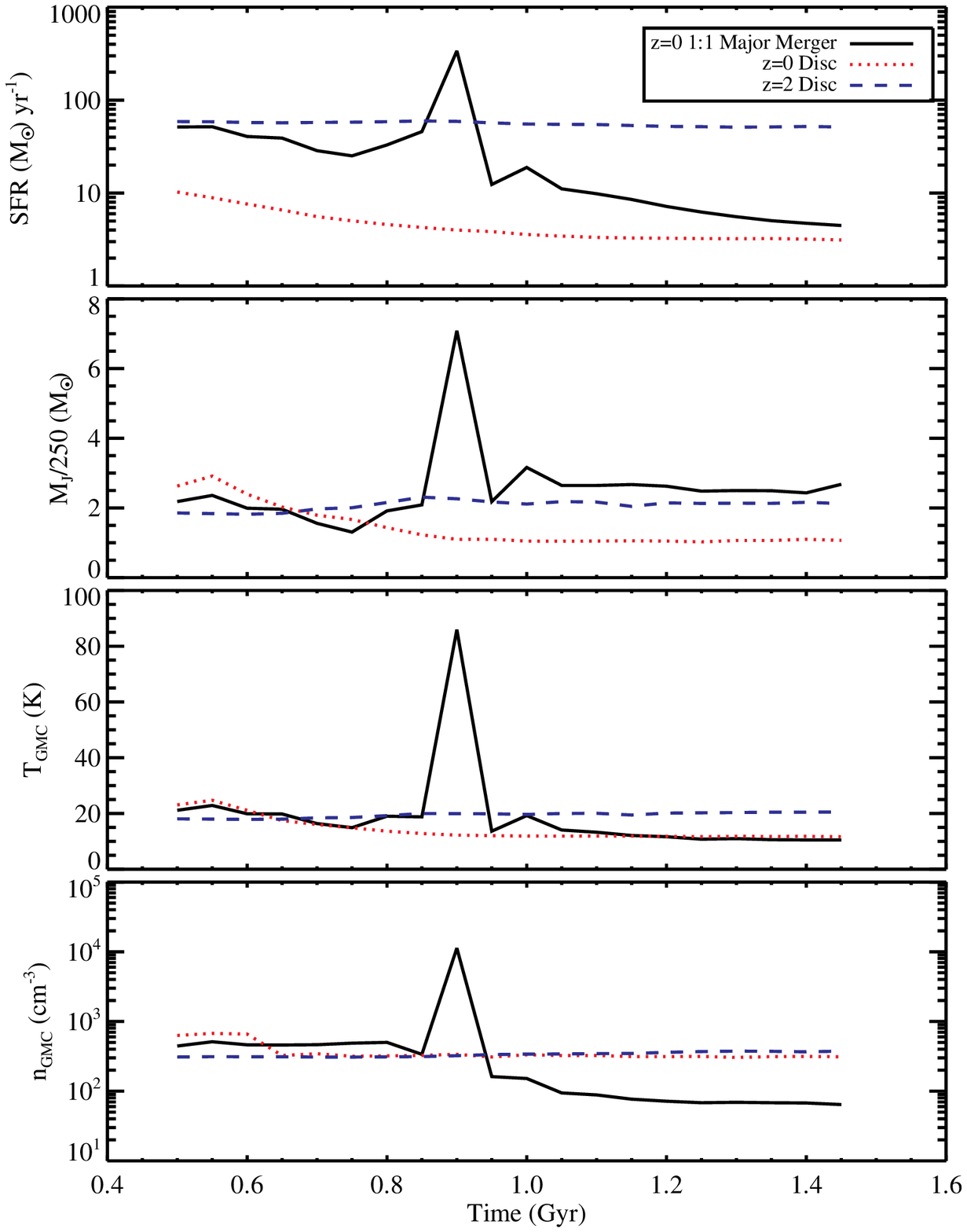}
\caption{Evolution of SFR, enhancement in \mjchar \ (above 250 \msun,
  a typical Jeans mass for an $n=10^2 \cmthree, T=10$ K GMC), \htwo
  \ mass-weighted GMC temperature and \htwo \ mass-weighted GMC density
  in a model \z=0 disc galaxy, \z=2 disc galaxy, and \z=0 1:1 major
  galaxy merger.  During the merger, warm dust heated by the starburst
  couples efficiently with the gas and raises the temperature of \htwo
  \ molecules.  This drives a rise in the typical Jeans mass of GMCs.
  \z=0 quiescent discs, on the other hand, have much more moderate
  conditions in their ISM. High-\z \ discs represent intermediate
  cases.\label{figure:mjevolution}}
\end{figure}

Because this study centres around understanding the Jeans scale in
GMCs, the thermal state of the molecular ISM must be known.  Following
\citet{kru11a,nar11b} and \citet{nar11c}, we consider the temperature
of the \htwo \ gas as a balance between heating by cosmic rays, grain
photoelectric effect, cooling by CO or CII line emission, and energy exchange
with dust.  The cosmic ray flux is assumed to be that of the mean
Galactic value\footnote{Tests described by \citet{nar11c} show that
  even under the assumption of a scaling where cosmic ray flux scales
  with the star formation rate of a galaxy, the typical thermal
  profile of \htwo \ gas in a given galaxy is unchanged.  This is
  because, in these environments, energy exchange with dust tends to
  dominate the temperature.}. Formally, if we denote heating processes
by $\Gamma$, cooling by $\Lambda$, and energy exchange with $\Psi$, we
solve the following equations:
\begin{eqnarray}
\Gamma_{\rm pe} + \Gamma_{\rm CR} - \Lambda_{\rm line} + \Psi_{\rm gd} = 0\\
\Gamma_{\rm dust} - \Lambda_{\rm dust} - \Psi_{\rm gd} = 0  
\end{eqnarray}
We refer the reader to \citet{kru11a} and \citet{nar11b} for the
equations regarding the photoelectric effect and cosmic ray heating
terms.

The dust temperature is calculated via the publicly available dust
radiative transfer code \sunrise \ \citep{jon06a,jon10a,jon10b}.  We
consider the the transfer of both stellar light from clusters of stars
as well as radiation from a central active galactic nucleus (though
this plays relatively little role in these simulations).  The stars
emit a \starburst \ spectrum \citep{lei99,vaz05}, where the metallicity
and ages of the stellar clusters are taken from the hydrodynamic
simulations.  The radiation then traverses the galaxy, being absorbed,
scattered and reemitted as it escapes.  The evolving dust mass is set
by assuming a constant dust to metals ratio comparable to the mean
Milky Way value \citep{dwe98,vla98,cal08}, and takes the form of the
$R=3.15$ \citet{wei01} grain model as updated by \citet{dra07}.  The
dust and radiation field are assumed to be in equilibrium, and the
dust temperatures are calculated iteratively.  Energy exchange between
gas and dust becomes important typically around $n\sim 10^{4}
\cmthree$.  Henceforth, we shall refer to this as the density where
grain-gas coupling becomes important.

  The gas cools either via CII or CO line emission.  The fraction of
  \htwo \ where most of the carbon is in the form of CO (versus atomic
  form) is determined following the semi-analytic model of
  \citet{wol10}, which is a metallicity dependent model (such that at
  lower metallicities, most of the carbon is in atomic form, and at
  higher metallicities, it is mostly in the form of CO): 
\begin{equation}
\label{eq:abundance}
f_{\rm CO} = f_{\rm H2} \times e^{-4(0.53-0.045 {\rm
    ln}\frac{G_0'}{n_{\rm H}/{\rm cm^{-3}}}-0.097 {\rm ln}Z')/A_{\rm v}}
\end{equation}
where $G_0'$ is the UV radiation field with respect to that of the
solar neighbourhood, and $A_{\rm V}$ is the visual extinction, $A_{\rm
  V} = \frac{N_{\rm H}}{1.87 \times 10^{21} {\rm cm^{-2}}}Z'$
\citep{wat11}.  If $f_{\rm CO} > 0.5$, we assume that the gas cools
via CO line emission; else, CII. The line emission is calculated via
the publicly available escape probability code as detailed in
\citet{kru07}.

While the temperature model is indeed somewhat complicated, the
typical temperature of a cloud can be thought of in terms of the
dominant heating effects at different densities.  At low densities
($n\sim 10-100 \ \cmthree$), the gas cools via line cooling to $\sim 8
$ K, the characteristic temperature imposed by cosmic ray heating.  At
high densities ($n\ga10^4 \ \cmthree$), grain-gas coupling becomes
efficient, and the gas rises to the dust temperature. At intermediate
densities, the temperature is typically in between $8 $ K and the dust
temperature.

\section{The Relation Between Jeans Mass and Star Formation Rate}
\label{section:imf}

We begin by examining the conditions under which the Jeans mass in
GMCs in a galaxy may vary.  We define the characteristic Jeans mass,
\mjchar, as the \htwo \ mass-weighted Jeans mass across all GMCs in a
galaxy, and consider the deviations of this quantity. The density term
in the Jeans mass calculation is the mean density of the GMC.  In
Figure~\ref{figure:mjevolution}, the top panel shows the SFR as a
function of time for an unperturbed \z=0 disc galaxy, a \z=0 major 1:1
galaxy merger, and a \z=2 disc.  The second panel shows the evolution
of $\mjchar/250~\msunend$.  250 \msun is the Jeans mass for physical
conditions as found in local discs like the Milky Way (MW), namely
$n=1-2 \times 10^2 \cmthree$ and $T=10 $ K.  Our disc galaxy has a
baryonic mass of $\sim 5\times10^{10}$ \msunend, and is the fiducial
MW model in \citet{nar11b}.  We note that the fiducial \z=0 MW model
naturally produces densities, temperatures, and \mjchar\ as observed
in the MW, thereby providing an important check for our ISM model.
The major merger is a merger of two $M_{\rm bar} = 1.5 \times 10^{11}$
\msun galaxies.  The high-\z \ disc model is an unperturbed $M_{\rm
  bar} = 10^{11} \msun$ disc.  The remaining panels show the evolution
of the characteristic temperatures and the characteristic GMC
densities\footnote{Note that because \mjchar\ is calculated as the
  \htwo\ mass-weighted Jeans mass, one cannot convert from the
  characteristic temperatures and densities shown in
  Figure~\ref{figure:mjevolution} to \mjchar.}.

For the bulk of the \z=0 disc's life, and during non-interacting
stages of the galaxy merger, the GMCs are relatively quiescent,
retaining surface densities near 100 \msunpcsq, and temperatures near
the floor established by cosmic ray heating of $\sim 10$ K.  During
the merger-induced starburst, however, the story is notably different.
Gas compressions caused by the nuclear inflow of gas during the merger
cause the average density to rise above $10^4 \cmthree$ in GMCs.  At
this density, grain-gas energy exchange becomes quite efficient.  At
the same time, dust heating by the massive stars formed in the
starburst causes the mass-weighted dust temperature to rise from $\sim
30$ K to $\sim 80$ K.  Consequently, the kinetic temperature of the
gas reaches similar values.  While the mean gas density also rises during
the merger, the Jeans mass goes as $T^{3/2}/n^{1/2}$.  As a result,
\mjchar \ rises by a factor $\sim 5$ during the merger-induced
starburst.  The \z=2 disc represents an intermediate case: high gas
fractions and densities drive large SFRs, and thus warmer conditions
in the molecular ISM. 

\begin{figure}
\hspace{-1cm}
\includegraphics[angle=90,scale=0.4]{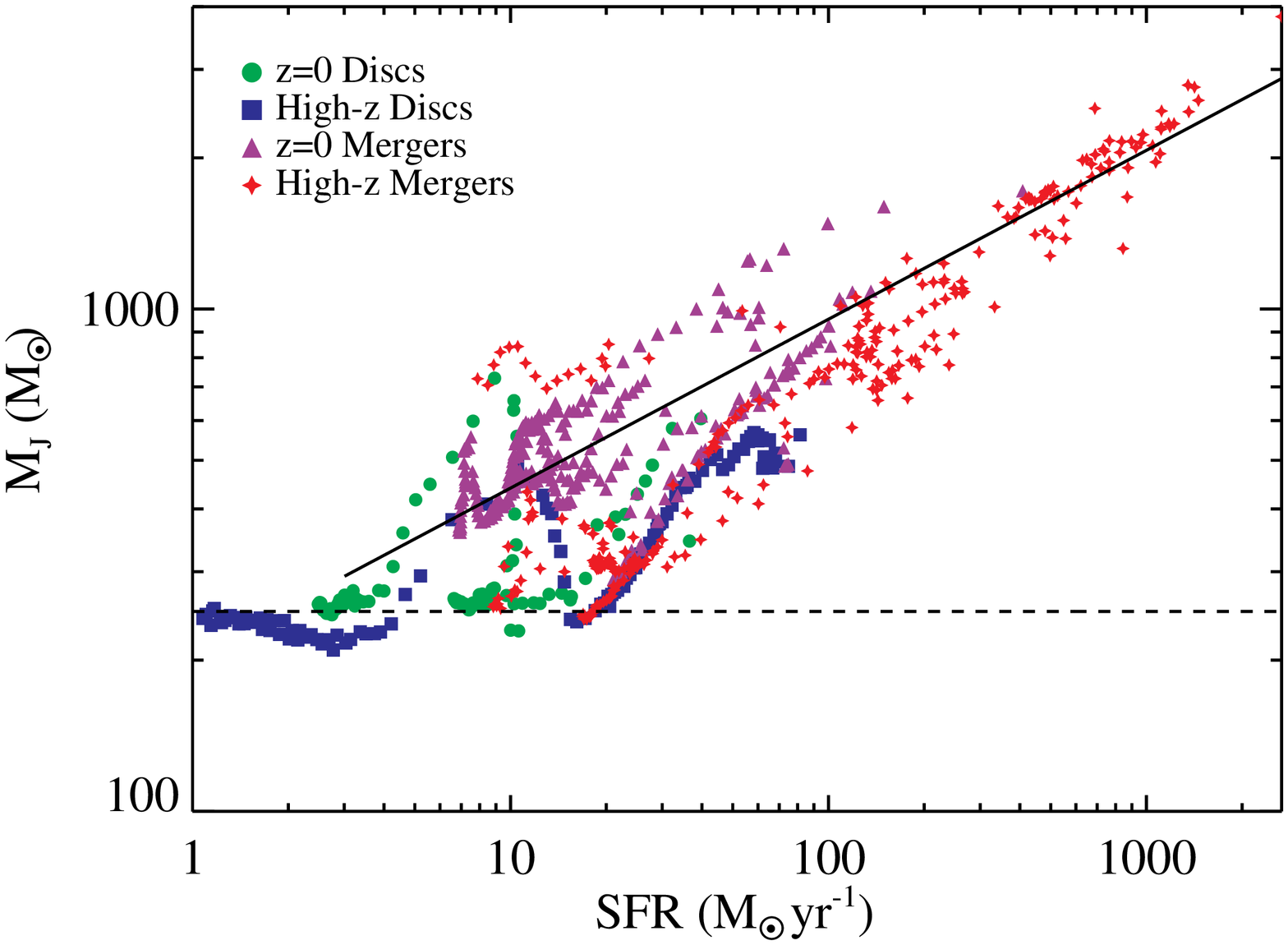}
\caption{Relationship between SFR and \htwo-mass weighted Jeans mass
  of GMCs for all simulation snapshots in our sample.  The points show
  individual simulation snapshots, and are colour-coded by the type of
  simulation.  The solid line denotes the best fit power law, while
  the dashed line shows a fiducial value of 250 \msunend, the typical
  Jeans mass for a GMC with density $n \approx 1-2 \times 10^2 \ 
  \cmthree$, and temperature $T \approx 10 $ K. Because dust and gas
  heating is dominated by star formation, $M_{\rm J}$  scales with
  the SFR. At low SFRs, the mean GMC density drops enough that
  gas-grain coupling becomes inefficient, and the heating is dominated
  by cosmic rays.  In this regime, the temperatures of the GMCs are
  roughly constant ($\sim 10 $ K), as are the Jeans
  masses. \label{figure:mjvsfr}}
\end{figure}

The characteristic Jeans mass in a galaxy is well-parameterised by the
SFR.  In Figure~\ref{figure:mjvsfr}, we plot the star formation rates
of all of the galaxies in our simulation sample against their
\mjchar. The points are individual time snapshots from the different
galaxy evolution simulations.  At high SFRs, dust is warmed by
increased radiation field, and thermally couples with the gas in dense
regions.  The increased temperature drives an increase in the Jeans
mass, resulting in roughly a power-law increase of \mjchar\ with SFR.

This relation does not extend to ${\rm SFR}\la3-5 \ \msunyrend$.  Here,
the mean densities become low enough ($n<10^4 \cmthree$) that energy
exchange with dust no longer keeps the gas warm.  Cosmic rays dominate
the heating in this regime, and the gas cools to $\sim 8-10 $ K, the
typical temperature that results from the balance of CO line cooling
and cosmic ray heating.  Because the temperatures are relatively
constant, and the characteristic densities of order $\sim 10-100
\ \cmthree$, the Jeans mass flattens with respect to SFR.  There may
be additional variations in \mjchar\ at smaller masses, owing e.g. to
metallicity variations that we neglect here.  For our purposes,
we are predominantly interested in higher-$z$ galaxies that generally
have SFR$\gg$5\msunyrend, so we do not try to characterise in detail
the behavior of \mjchar\ at lower SFR.

The resulting relation is well-fit by a powerlaw $\mjchar/250 \msun =
0.5\times {\rm SFR}^{0.4}$ at ${\rm SFR}\ga 3 \msunyrend$.  This fit
is shown in Figure~\ref{figure:mjvsfr} by the black line, and extends up
to $\sim 1000 \ \msun/$yr in our models.  The fit excludes the low SFR
tail at SFR $< 3 \ \msunyrend$.  There is substantial scatter around
the relation owing to the fact that at a given SFR, there may be a
variety of physical conditions.  For example, galaxies with SFR $\sim
50 \ \msunyrend$ may represent merger-induced starbursts with extremely
warm nuclear star formation, as well as high-\z \ discs with more
distributed star formation and moderate temperatures.

We choose to parameterise \mjchar \ in terms of the SFR in order to
facilitate analysis of inferred observed SFRs of high-\z \ galaxies.
In principle, one could imagine a relationship between \mjchar \ and
stellar mass.  Indeed, broadly, the lower stellar mass galaxies tend
to have low \mjchar, and vice versa.  However, due to similar stellar
masses on (e.g.) before and after a starburst event, the scatter in an
\mjchar-$M_*$ relation is even larger.  Similarly, \mjchar \ may
correlate well with the \htwo \ gas surface density.  Indeed,
\citet{nar11c} showed that the physical conditions in the gas
(temperature and velocity dispersion) scale well with the molecular gas
surface density.  The scatter in such a relation, however, is similar
to the \mjchar-SFR relation \citep[the SFR is forced to scale with
  the volumetric gas density, and shows a tight correspondence with
  the gas surface density for these simulations;][]{nar11a}.

One could similarly imagine a Jeans mass that is determined by the
median density in the gas, rather than the mean density.  As it turns
out, this has little effect on our results. We discuss this issue in
further detail in Appendix~\ref{section:appendix}.

Finally, we emphasise that what governs the Jeans mass in GMCs is the
physical conditions in the ISM, not the global morphology.  As such,
both heavily star-forming discs (e.g. the \bzk \ population at \zsim
2) as well as galaxy mergers may exhibit warmer ISM gas relative to
today's discs.  One can expect short-lived nuclear starbursts driven
by major mergers in the local Universe to exhibit an increase in
\mjchar \ by a factor of $\ga 5$.  Similarly, high-\z\ disc galaxies
that are forming stars at many tens of \msunyr\ for a much longer duty
cycle may show an increased \mjchar\ in their GMCs by a factor $\sim
2-3$.  The increase in temperature of the ISM is typically driven by
warm dust and increasing fractions of molecular gas above the
gas-grain coupling density with increasing SFR.  We note that very
high redshifts, when comparing to galaxies at a fixed density, then
the Jeans mass will only scale with the SFR once the mean gas
temperature is warmer than CMB temperature.  We discuss this more
quantitatively in \S~\ref{section:madau}.

\section{Implications for Observed Star Formation Rates}
\label{section:sfr}

We showed that the \htwo\ mass-weighted Jeans mass can vary substantially
in galaxies with physical conditions different than that in the Milky Way.
We now employ make the assumption that variations in \mjchar\ directly
into variations in the characteristic mass \mc\ of the IMF.  This then
allows us to investigate how such variations in \mjchar\ will change
the inferred star formation rates.

It is useful to parameterise this variation in terms of the how the
bolometric luminosity \lbol\ varies with SFR.  In a star-forming
galaxy, \lbol\ is essentially dominated by the emission from high
mass stars.  All currently used extragalactic SFR tracers trace
such stars.  Hence examining the variations in \lbol\ is likely to
be broadly representative of the variations that would be seen in
any of the canonical SFR tracers.  For an invariant IMF, \lbol\ is
linearly related to SFR (assuming, as we will throughout, that there
is no significant contribution from active galactic nuclei).  But
when the IMF varies, then \lbol\ will trace only high-mass stars,
while the true SFR includes all stars, thereby breaking the linear
relation.

To quantify the SFR-\lbol\ relation, we begin by assuming that the
IMF has a broken power-law form characterised by a turnover mass,
\mc, a low-mass slope of 0.3, and high-mass slope of -1.3 (where
the IMF slope specifies the slope of the log$_{\rm 10}$(dN/dlogM)-log$_{\rm
10}$(M) relation).  If \mc\ scales with $M_{\rm J}$, then we can
calculate the variation in the \lbol-SFR relationship as a function
of \mc\ by employing stellar population synthesis modeling. 

We employ \fsps, a publicly available stellar population synthesis
code \citep{con09b,con10a,con10b}.  \fsps\ allows for the quick
generation of stellar spectra given a flexible input form for the IMF.
We assume a constant star formation history (SFH) of 10~\msunyrend.
This was chosen as a reasonable approximation for the star formation
history of a (\z=0) $M_*=10^{10.5} \msun$ galaxy in cosmological
simulations \citep{dav08}.  We additionally assume solar metallicity
stars, and include the contribution from thermally pulsating
asymptotic giant branch (AGB) stars.  No dust extinction is included.

Using \fsps, we find that \lbol/SFR $\sim \mc^{0.3}$.  To test our
model assumptions, we vary the assumed star formation history and
assume a constant SFH of 50 \msunyrend, as well as an exponentially
declining SFH with an initial SFR of 50 and 10 \msunyr and an
$e$-folding time of 10 Gyr. All assumed SFH models return an
\lbol/SFR-\mc \ exponent within 10\% of our fiducial model.  Combining
our fiducial \lbol/SFR-\mc \ relation with the model fit to the
SFR-\mc \ relation fit from Figure~\ref{figure:mjvsfr}, we arrive at
the relation:
\begin{equation}
\label{eq:sfrlir}
{\rm SFR} = \left[\frac{\lbol}{10^{10}\lsunend}\right]^{0.88} \msunyr, \;\;\; {\rm for\; SFR}\ga 3\msunyrend
\end{equation}
As seen in Figure~\ref{figure:mjvsfr}, at ${\rm SFR} \la 3 \msunyrend$
there is no strong variation of \mjchar\ with SFR, so in that regime
SFR$\propto\lbol$.

  At ULIRG-type SFRs ($\sim 100 \ \msunyrend$), our model suggests
  that SFRs may need to be adjusted by a factor $\sim 2$.  At the
  luminosities of distant SMGs, the most luminous, heavily
  star-forming galaxies in the Universe, typical SFRs may be closer to
  $\sim400\ \msunyrend$, rather than the canonically inferred $\sim
  1000\ \msunyrend$.


Though the correction to the inferred star formation rates from a
standard Kroupa IMF is relatively mild (only a factor of $\sim 3$ at a
SFR of $\sim 1000 \ \msunyrend$), the impact on cosmological star
formation rates can be substantial owing to the strong evolution of
global SFRs.  In the following subsections, we discuss the impact of a
SFR-\lbol \ relation that derives from a variable IMF on the observed
Kennicutt-Schmidt star formation relation, the SFR-$M_*$ relation,
high-redshift sub-millimetre galaxies, and the observed cosmic star
formation rate density.

\subsection{The Kennicutt-Schmidt Star Formation Law}
\label{section:ks}

\begin{figure}
\includegraphics[angle=90,scale=0.4]{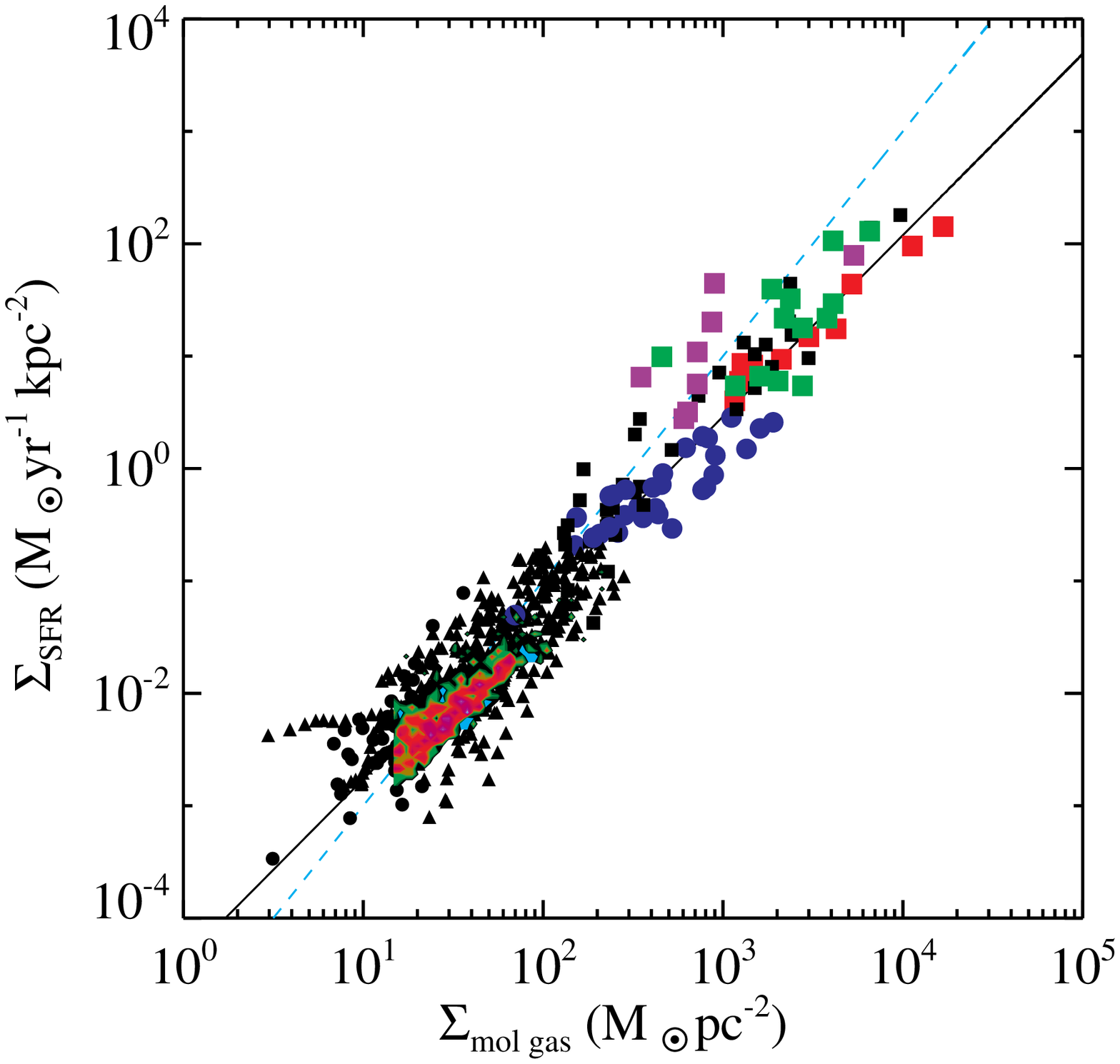}
\caption{Kennicutt-Schmidt $\Sigma_{\rm SFR}-\Sigma_{\rm mol}$ star
  formation law. Solid line shows relation $\Sigma_{\rm SFR} \sim
  \Sigma_{\rm mol}^{1.6}$ relation that results when utilising the SFR
  corrections derived for the model IMF in this paper, and the model
  functional form for the CO-\htwo \ conversion factor presented in
  \citet{nar11c}.  The blue dashed line denotes the relation
  $\Sigma_{\rm SFR} \sim \Sigma_{\rm mol}^2$, which results if the IMF
  does not vary with the Jeans mass of GMCs.  The low-\z \ data are
  from \citet{ken07,won02,cro07,sch07,big08,ken98b}, whereas the
  high-\z \ data are from
  \citet{gen10,dad10b,dad10a,bot09,bou07,gre05,tac06,tac08,eng10}.\label{figure:ks}}
\end{figure}

The Kennicutt-Schmidt (KS) star formation relation is a power-law
relationship between the star formation rate surface density against
the gas surface density of both sub-regions of galaxies, as well as
galaxies as a whole.  A reasonably tight relationship has been
observed on scales as small as GMC clumps
\citep{wu05,wu10,hei10,lad11}, $\sim 10^2-10^3 $ pc scale regions in
galaxies \citep[e.g.][]{ken98b,big08,sch11}, and galaxies as a whole both at
low and high-\z
\ \citep[e.g.][]{gao04a,gao04b,nar05,bus08,ion09,dad10b,gen10,gar11}. The
relationship between the star formation rate and gas density is
crucial both for our understanding of the dominant physical effects in
controlling star formation rates \citep[e.g.][to name just a
  few]{sil97,elm02a,kru05,tan10,ost11}, as well as for input into
galaxy evolution simulations \citep[e.g.][]{spr03a,rob08,bou10,cev10}.


The KS relation is typically measured observationally via a
combination of inferred star formation rates, and measured \htwo \ gas
masses.  At face value, assuming a constant \lbol-SFR relation, and a
constant CO-\htwo \ conversion factor, a relation $\Sigma_{\rm
  SFR}-\Sigma_{\rm H2}^{1.3-1.5}$ results when including both local
and high-\z \ data \citep[e.g. ][]{dad10b,gen10}\footnote{We note that
  when considering only local data from quiescent discs, the KS
  relation has a roughly linear index \citep{big08}.  It is only when
  including higher surface-density galaxies such as ULIRGs or high-\z
  \ galaxies that the fit is torqued toward higher values
  \citep{kru09b}.}.  However, it is now well established that the
CO-\htwo\ conversion factor is not universal.  Observations
\citep[e.g.][]{bol08,tac08,ler11,gen11b}, as well as theory
\citep{she11a,she11b,fel11,nar11b,nar11c} motivate a continuous form
of \xco \ in terms of the physical properties of the \htwo \ gas.
\citet{nar11c} derived a functional form for the CO-\htwo\ conversion
factor (\xco) as a function of CO surface brightness as an observable
proxy for the gas density and velocity dispersion.  These authors
showed that combining their model of \xco\ with literature data
results in a steeper inferred KS relation, $\Sigma_{\rm SFR} \sim
\Sigma_{\rm H2}^2$.  This owes to decreasing CO-\htwo\ conversion
factors at higher SFR surface densities, and assumes a linear
\lbol-SFR relation.

If the stellar IMF varies with the physical conditions in the ISM, the
KS relation will be further modified, since \lbol-SFR is no longer
linear. Equation~\ref{eq:sfrlir} suggests when utilising a standard
Kroupa IMF, the inferred SFR for the most heavily star-forming
galaxies may be overestimated, and that the KS power-law index will
decrease.  In Figure~\ref{figure:ks}, we plot the compiled data from
\citet{nar11c} which includes both resolved regions from nearby
galaxies, as well as global data from quiescent discs and starburst
galaxies from $\z\approx0-2$, on a KS relation.  The data have had
their SFR's calculated via Equation~\ref{eq:sfrlir}.  The best fit
relation is $\Sigma_{\rm SFR} \sim \Sigma_{\rm H2}^{1.6}$.  For
reference, we show the \citet{nar11c} relation, $\Sigma_{\rm SFR} \sim
\Sigma_{\rm H2}^2$ relation via the blue dashed line.

In order to calculate the modified SFRs in Figure~\ref{figure:ks}, we
assume that the original SFRs reported in the papers that the data
originate from are relateable to the \lir\ via the standard
\citet{ken98a} relation, and that $\lbol \approx \lir$.  This is
likely a reasonable assumption.  Galaxies below SFR of 3 \msunyr do
not undergo a modification to the SFR in our model. The galaxies which
will have the strongest modification to their SFR when utilising our
model IMF are those that form stars at the rates of SFR $\ga 100
\ \msunyrend$.  These galaxies are typically mergers in the local
Universe, and either heavily star-forming discs or mergers at high-\z,
and the bulk of their luminosity is in the infrared.

Any theory that relates the SFR to the free-fall time in a GMC
naturally results in a relation SFR $\sim \rho_{\rm H2}^{1.5}$.  For
galaxies of a constant scale height, this translates into a similar
surface density relation.  An assumption of a constant scale height
across such a wide range of physical conditions is debatable, however
\citep{she08,kru12a}.  Alternatively, if SFR $\sim \rho$ for quiescent
regions, and has a steeper relation for starburst galaxies as expected
for a scenario in which gas dominates the vertical gravity
\citep{ost11}, then a surface density relation steeper then unity is
natural.


\subsection{The Dynamical Time Star Formation Law}
\label{section:silk}

\begin{figure}
\includegraphics[angle=90,scale=0.4]{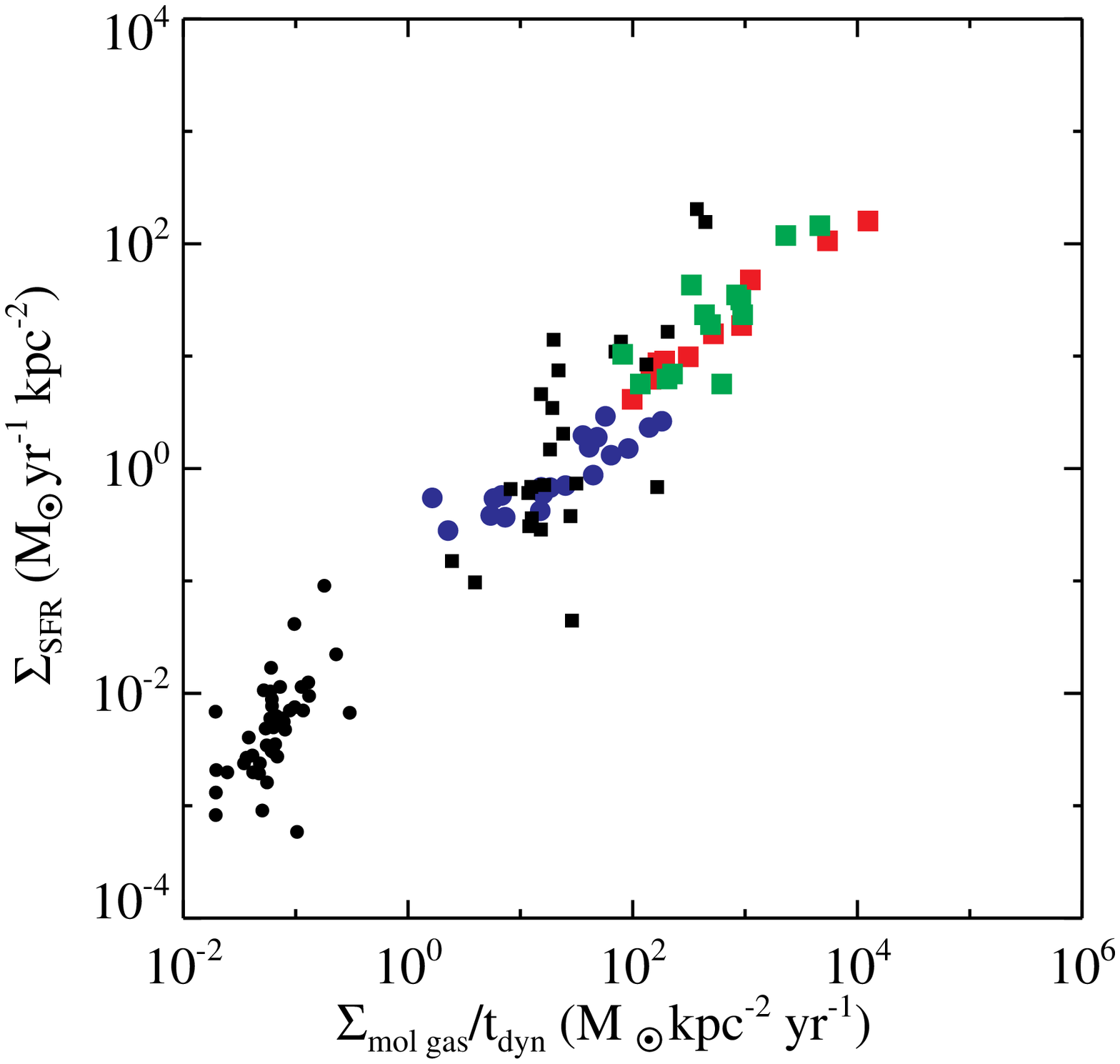}
\includegraphics[angle=90,scale=0.4]{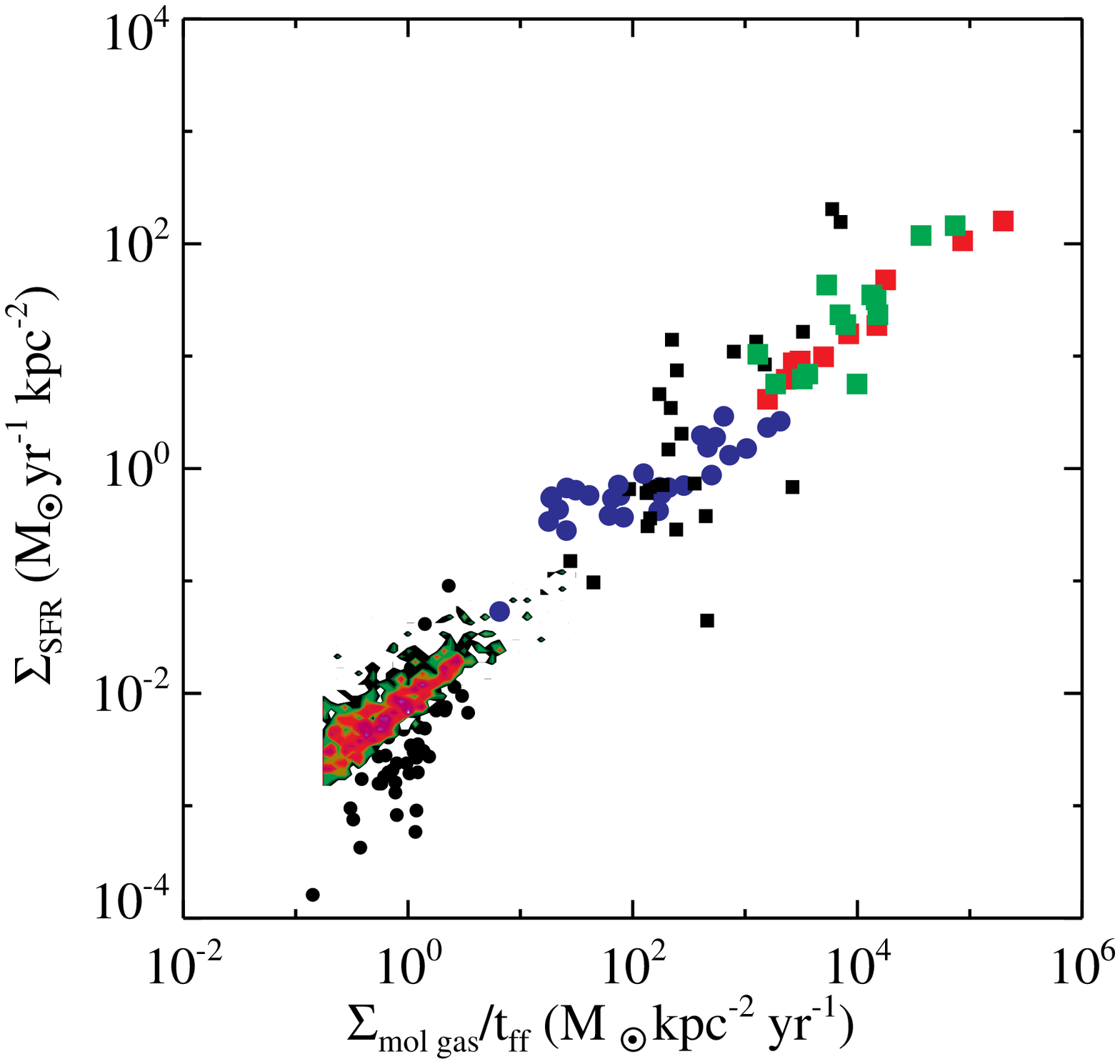}
\caption{Similar to Figure~\ref{figure:ks}, but with gas surface
  densities divided by the dynamical time (top) or free-fall time
  (bottom).  Both relations are nearly linear, with star formation
  efficiencies $\epsilon_{\rm ff} \sim 0.01$. See text for
  details. \label{figure:ks_silk}}
\end{figure}

As an alternative to the standard KS relation that relates
$\Sigma_{\rm SFR}$ to $\Sigma_{\rm H2}$, some theories posit that the
gas density divided by the galaxy dynamical time or local free fall
time may be the relevant physical parameter controlling the SFR
\citep[e.g.  ][]{sil97,elm02a}.  It is of interest, therefore, to
estimate the modified $\Sigma_{\rm SFR}-\Sigma_{\rm H2}/t_{\rm dyn}$
and $\Sigma_{\rm SFR}-\Sigma_{\rm H2}/t_{\rm ff}$ relations given our
model IMF which varies with the physical conditions in the ISM.  We
plot these relations in Figure~\ref{figure:ks_silk}.

Following \citet{dad10b} and \citet{gen10}, we define the dynamical
time of the galaxies in Figure~\ref{figure:ks} as the rotational
time either at the galaxy's outermost observed radius, or the half
light radius, depending on the sample.  When assuming a constant
SFR-\lbol\ relation\footnote{Note, that this relation depends on
the usage of a physically motivated functional form for the CO-\htwo\
conversion factor that depends on the CO surface brightness and the
metallicity of the galaxy \citep{nar11c}.}, $\Sigma_{\rm SFR}$ is
nearly linearly related to $\Sigma_{\rm mol}/t_{\rm dyn}$, with
exponent $\sim 1.03$ \citep{nar11c}.  When we use Equation~\ref{eq:sfrlir}
to calculate the star formation rates, a relation $\Sigma_{\rm SFR}
\sim \Sigma_{\rm mol}/t_{\rm dyn}^{0.9}$ results.  Hence this
IMF-driven variation produces only a minor change to the relation
relative to a constant SFR-\lbol\ relation.  

Why does this relation change only marginally when the change in
the $\Sigma_{\rm SFR}-\Sigma_{\rm mol}$ was more pronounced?
$\Sigma_{\rm mol}/t_{\rm dyn}$ spans $\sim 6$ orders of magnitude
when considering the galaxies in Figure~\ref{figure:ks}, whereas
$\Sigma_{\rm mol}$ spans only $\sim 4$.  The drop in the SFR of the
most heavily star forming galaxies when assuming our model IMF makes
less of an impact on the inferred relation than in the $\Sigma_{\rm
SFR}-\Sigma_{\rm mol}$ case.

Similarly, we can consider how our model affects the $\Sigma_{\rm
SFR}-\Sigma_{\rm mol}/t_{\rm ff}$ relation.  Both observations and
models have suggested that the star formation efficiency per free
fall time should be roughly constant in molecular gas
\citep{kru05,kru07b,eva09,ost11,kru11c}.  Following \citet{kru11c},
we infer the free fall time from the observable properties of the
galaxy.  Similar to the $\Sigma_{\rm SFR}-\Sigma_{\rm mol}/t_{\rm
dyn}$ relation, when assuming a constant SFR-\lbol \ relation,
$\Sigma_{\rm SFR}$ varies roughly linearly with $\Sigma_{\rm
mol}/t_{\rm ff}$ \citep{nar11c}.  Introducing a variable IMF in the
SFR calculations results in a moderate change the $\Sigma_{\rm
SFR}-\Sigma_{\rm mol}/t_{\rm ff}$ exponent to $\sim 0.9$ for similar
reasons as in the $t_{\rm dyn}$ calculation.  A star formation
efficiency $\epsilon_{\rm ff} \approx 0.01$ provides a good fit to
the relation $\Sigma_{\rm SFR} = \epsilon_{\rm ff}\Sigma_{\rm
mol}/t_{\rm ff}$.

\begin{figure*}
\includegraphics[angle=90,scale=0.7]{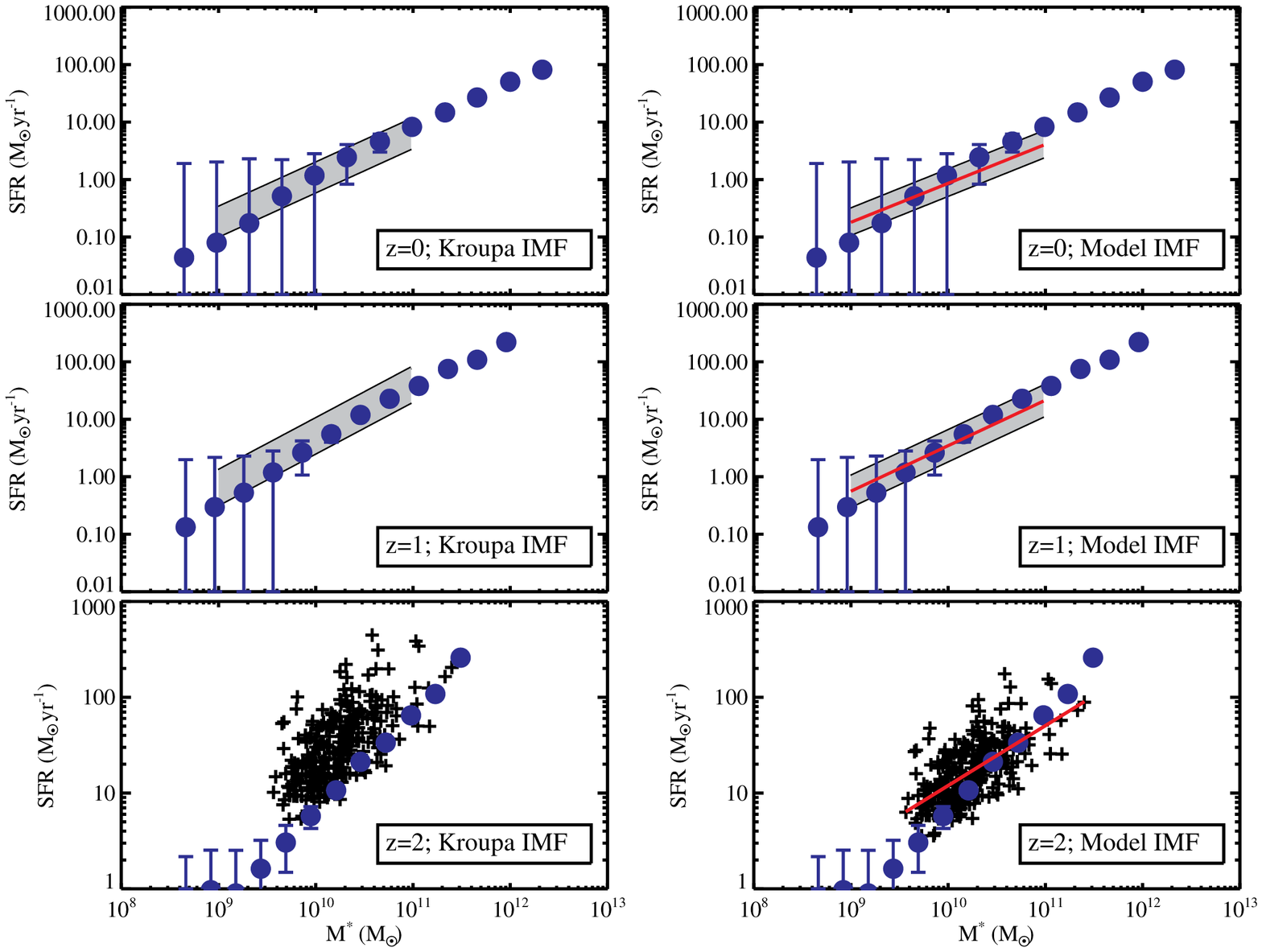}
\caption{SFR-$M_*$ relationship for \z=0,1,2 galaxies.  Left columns
  are literature values of SFR, and right show corrected SFRs using
  our model IMF.  In the top two rows, the shaded grey region shows
  the median observed relationship from \citet{elb07}, with
  dispersion, and in the bottom row the crosses come from
  \citet{dad07}.  In all panels, the circles with error bars show the
  simulated points from \citet{dav11b} in bins of $M_*$, with
  dispersion.  The red lines in the right column show the best fit to
  the modified SFR-$M_*$ relation.  When using a standard Kroupa IMF,
  the simulated galaxies underpredict the SFRs by as much as a factor
  of $\sim 10$.  When assuming an IMF that scales with the thermal
  conditions in the molecular ISM, the simulated SFR-$M_*$ relation
  and observed one comes into much better agreement.  This is most
  apparent in the \z=2 case (bottom row). \label{figure:sfr_mstar}}
\end{figure*}

\subsection{The SFR-$M_*$ Relation}
\label{section:sfr-mstar}

The stellar masses and star formation rates of galaxies have been
observed to have a relatively tight relationship as early as \z=6-8
\citep{noe07b,noe07a,dad07,sta09,mcl11}.  The SFR-$M_*$ relation
appears to steadily decrease in normalisation with time, in accordance
with cosmological simulations that suggest that star formation is
fueled by gravitational infall of baryons into the
galaxy~\citep[e.g.][]{ker05,dav11b}, which shows a rapid drop with
redshift~\citep[e.g.][]{dek09}.

As alluded to in \S~\ref{section:introduction}, while simulations
predict the slope and scatter of the SFR-$M_*$ relation with
reasonable success, the amplitude at $\z\approx 2$ is less well
reproduced.  In models, the SFRs at a given $M_*$ at $z=2$ are too
low by $\times2-3$ compared with observations.  This is a fairly
small discrepancy given the systematic uncertainties, but is
nonetheless persistent for simulations performed via a variety of
techniques (hydrodynamic simulations performed using SPH or adaptive
mesh refinement, as well as SAMs).

This discrepancy led \citet{dav08} to suggest that perhaps the SFRs
of \zsim 2 galaxies are overestimated.  Dav\'e empirically derived
an IMF that varies its critical mass with redshift as $\mc =
0.5(1+\z)^2 \ \msun$ for $0<z<2$ as being able to reconcile the
models and data.  Our model for a varying IMF differs from the
\citet{dav08} model in that here, the IMF varies with the temperature
and density of GMCs in the ISM, rather than with redshift.  Thus,
a heavily star-forming galaxy in the local Universe may have an IMF
more comparable to a luminous disc at \zsim 2 than a present-epoch
quiescent disc.  Nevertheless, the impact of such a varying IMF
does imply a strong dependence in redshift because high-redshift
galaxies form stars far more rapidly.

In Figure~\ref{figure:sfr_mstar}, we plot the observed SFR-$M_*$
data from the Sloan Digital Sky Survey (SDSS; \z=0) and Great
Observatories Origins Deep Survey (GOODS; \z=1-2), overlaid by the
predicted SFR-$M_*$ relation from simulated galaxies drawn from the
hydrodynamic cosmological simulations of \citet{dav11b}.
The left column shows the observed and modeled SFR-$M_*$ relation
assuming a Kroupa IMF with turnover mass \mc=0.5 \msunend.  With
increasing redshift, the discrepancy between the normalisation of the
observed SFR-$M_*$ relation and modeled one grows, becoming noticeable
at $z\ga 1$. 

If the turnover mass in the IMF varies with the Jeans mass in GMCs,
this discrepancy may be ameliorated.  Turning now to the right
column, we show the same comparison after correcting the SFRs via
Equation~\ref{eq:sfrlir}.  While the relatively quiescent galaxies
in the observed low-\z\ samples undergo relatively little correction
to their already modest SFRs, the galaxies at high-\z\ tend to be
brighter at a given stellar mass than their low-\z\ counterparts
due to increased gas fractions and densities.  The SFRs for these
galaxies undergo a larger correction per Equation~\ref{eq:sfrlir},
and bring the observed SFR-$M_*$ relation into agreement with
simulations.

We fit a power-law form to observations, as reinterpreted using our
evolving IMF.  For the \z=0 and \z=1 data, we fit the change in the
median relations reported by \citet{elb07}.  For the \z=2 data, we
perform a polynomial fit to the data from \citet{dad07}\footnote{Note
  that we do not have error bars available for the \citet{dad07}
  data.}.  Our best fit relations are:
\begin{eqnarray}
(\z=0): {\rm SFR} (\msunyrend) = 3.9 \times \left[\frac{M_*}{10^{11} \msun}\right]^{0.65}\\
(\z=1): {\rm SFR} (\msunyrend) = 19.3 \times \left[\frac{M_*}{10^{11} \msun}\right]^{0.77}\\
(\z=2): {\rm SFR} (\msunyrend) = 50.2 \times \left[\frac{M_*}{10^{11} \msun} \right]^{0.55}
\end{eqnarray}
and are denoted in Figure~\ref{figure:sfr_mstar} by the solid red
lines in the right column. 

It important to note that all of these relations should be regarded
as tentative -- they represent relatively limited ranges of stellar
masses, and for the \z=0 and \z=1 data the relations were derived
by re-fitting published fits, not the actual data with error bars.
This is therefore only intended to be illustrative of the sense and
magnitude of the changes in SFR-$M_*$ when applying an IMF with
$\mc\propto \mjchar$.  The best fit exponents for \z=[0,1,2] as
reported by \citet{dad07} and \citet{elb07} are $\sim$[0.77,0.9,0.9]
respectively, which in our scenario decrease to $\sim$[0.65,0.77,0.55].
This of course owes to the fact that in our model, galaxies with
larger inferred SFRs (using a standard Salpeter or Kroupa IMF)
undergo a larger correction to their SFR when the IMF varies with
the Jeans mass.  Similarly,
the amplitudes of the SFR-$M_*$ relations in Figure~\ref{figure:sfr_mstar}
are lower than previously estimated due to lower SFRs.

The cosmological simulation predictions now match the observations
quite well in amplitude (around $\sim L^*$), but the slope is now more
discrepant.  A selection effect that could explain this discrepancy is
that low SFR galaxies tend to be absent from these samples, creating a
floor in observed SFRs that increases to higher redshifts.  Combined
with the (not insubstantial) scatter in the SFR-$M_*$ relation, this
could induce a shallower observed slope relative to the true slope
(N. Reddy et al., in prep.).  On the other hand, it could also just
indicate a failing in these simulations.  We leave a more detailed
comparison to models for a later time (since in detail, an evolving
IMF would have secondary implications for galaxy evolution in these
models that are not accounted for by this simple rescaling).  For now,
we simply illustrate how galaxy SFRs could be significantly lowered by
an evolving IMF during the peak epoch of cosmic star formation.

 An important constraint on SFR-$M_*$ evolution is that its amplitude
 appears to be essentially constant from $z\sim
 2-6$~\citep{sta09,bou11b}.  Simulations, on the other hand, tend to
 predict an amplitude that continues to rise, albeit at a slower pace
 than at lower redshifts.  Hence as discussed in \citet{dav11c}, if
 the IMF varies in some systematic way with redshift in order to
 reconcile models and data, then the evolution must {\it reverse} at
 higher redshifts.  This would require that the typical SFRs of
 galaxies at $z\ga 2$ should be lower at earlier epochs.  Such an
 evolution is at least qualitatively consistent with the observation
 that $L^*$ decreases going to higher redshifts (i.e. the Universe is
 ``upsizing" at that early epoch, not downsizing).  On the other hand,
 at very high redshifts, the CMB temperature rises accordingly,
 setting a larger minimum ISM temperature (and hence a larger minimum
 Jeans mass).  In this sense, it is not quite obvious how to reconcile
 the potential discrepancy between the observed normalisation of the
 SFR-$M_*$ relation and modeled one.  A quantitative comparison will
 require a more thorough analysis of the data versus the models, which
 we leave for the future.


\subsection{Sub-millimetre galaxies}
\label{section:smg}

Sub-millimetre galaxies at high redshifts are the most rapidly
star-forming galaxies known, and hence likely represent an extreme
of ISM physical conditions where our IMF scenario would predict the
largest variations.  Indeed, theoretical models attempting to
reproduce SMGs have sometimes resorted to invoking IMF variations
in order to explain these systems.

Semi-analytic models with a canonical IMF have difficulty matching
the number counts of high-\z \ infrared and Sub-millimetre luminous
galaxies \citep{bau05,gon11,nie12}, and have invoked as a solution
that the IMF is more top-heavy in these merger-driven starbursting
systems.  On the other hand, the required IMF variation in this
scenario is rather extreme, namely a flat IMF above $1 M_\odot$,
which is disfavoured by observational constraints on the IMF from
measuring the gas, stellar, and dynamical mass of SMGs~\citep{tac08}.

The simulations of \citet{dav10} suggest a different origin for SMGs,
as the high-mass, high-SFR end of the $z\sim 2-3$ galaxy main
sequence.  While these models can broadly reproduce many properties of
SMGs, they do not achieve the high SFRs inferred from the SMG
observations, falling short by a factor of $\sim3$.  This is similar
to the correction factor predicted in Equation~\ref{eq:sfrlir} for the
most IR luminous galaxies.  Hence if the IMF varied in the manner
considered here, this scenario for the origin of SMGs would be
strengthened.

Conversely, recent hydrodynamic simulations of isolated galaxies and
galaxy mergers (similar to those presented here), coupled with dust
radiative transfer calculations and semi-empirical cosmological
simulations have shown that the number counts of
high-\z\ sub-millimetre selected galaxies may be accounted for while
utilising a Kroupa IMF \citep[][]{nar10a,hay10,hay11}, at least within
the current uncertainties.  These models suggest that pairs of
galaxies blended within a single far-IR beam may be an important
observational bias at lower flux densities, and that the relationship
between sub-millimetre flux density and SFR may not be linear owing to
substantial variations in dust temperatures during the merger phase.
If the observed number counts from upcoming surveys turn out to be in
good agreement with these predictions at a range of far-IR
wavelengths, then an IMF-based solution may be unnecessary.

While they differ in detail, both the \citet{dav10} and \citet{hay11}
models suggest that the lowest luminosity SMGs ($\sef \sim 5 $ mJy)
galaxies may represent individual, heavily star-forming discs, whereas
more luminous SMGs likely owe their origin to galaxy mergers.  In
hydrodynamic cosmological simulations, individual discs rarely sustain
SFRs at rates larger than $\sim 500 \ \msunyrend$
\citep{dek09,dav10,hay11}.  Utilising Equation~\ref{eq:sfrlir}, this
corresponds to luminosities $\lbol \approx 10^{13}$ \lsun, and may
reflect a transition luminosity where objects above this are typically
mergers at \zsim 2.  Recent analysis of the specific star formation
rates and \lir-$T_{\rm dust}$ relation for high-\z \ SMGs by
\citet{mag12} suggests that this may indeed represent a transition
luminosity between galaxy populations.  At $\lbol \approx 10^{13}$
\lsun, the \lir-$T_{\rm dust}$ relation steepens dramatically, and
galaxies rise above the ``main-sequence'' of specific star formation
rates at \zsim 2 \citep{dad07,rod11}.  Moreover, models by
\citet{hop10} show that $\lbol \approx 10^{13} \lsun$ represents a
transition luminosity above which the luminosity function is dominated
by mergers at \zsim 2.  If a Salpeter IMF holds for \zsim 2 SMGs, a
SFR of 500 \msunyr would correspond to $\sim 3 \times 10^{12} \lsun$,
an implausibly low transition luminosity given \zsim 2 merger rates
\citep{fak08,guo08,hop10}.


In short, achieving the extreme SFRs inferred for SMGs at their
observed number densities using a canonical IMF remains a challenge
for hierarchical-based models.  While observations are sufficiently
uncertain to be broadly consistent with a variety of scenarios,
tensions could generally be alleviated if the SFRs were a factor of
several lower in these systems as would be the case for the varying
IMF presented here.  Our model form for the IMF suggests a reasonable
transition luminosity for the break between discs and mergers at \zsim
2, given a rough maximal SFR for a quiescent disc of $\sim 500
\ \msunyrend$ ($\lbol \approx 10^{13} \lsun$).

\subsection{The Evolution of Cosmic Star Formation Rate Density}
\label{section:madau}

\begin{figure}
\hspace{-1cm}
\includegraphics[angle=90,scale=0.4]{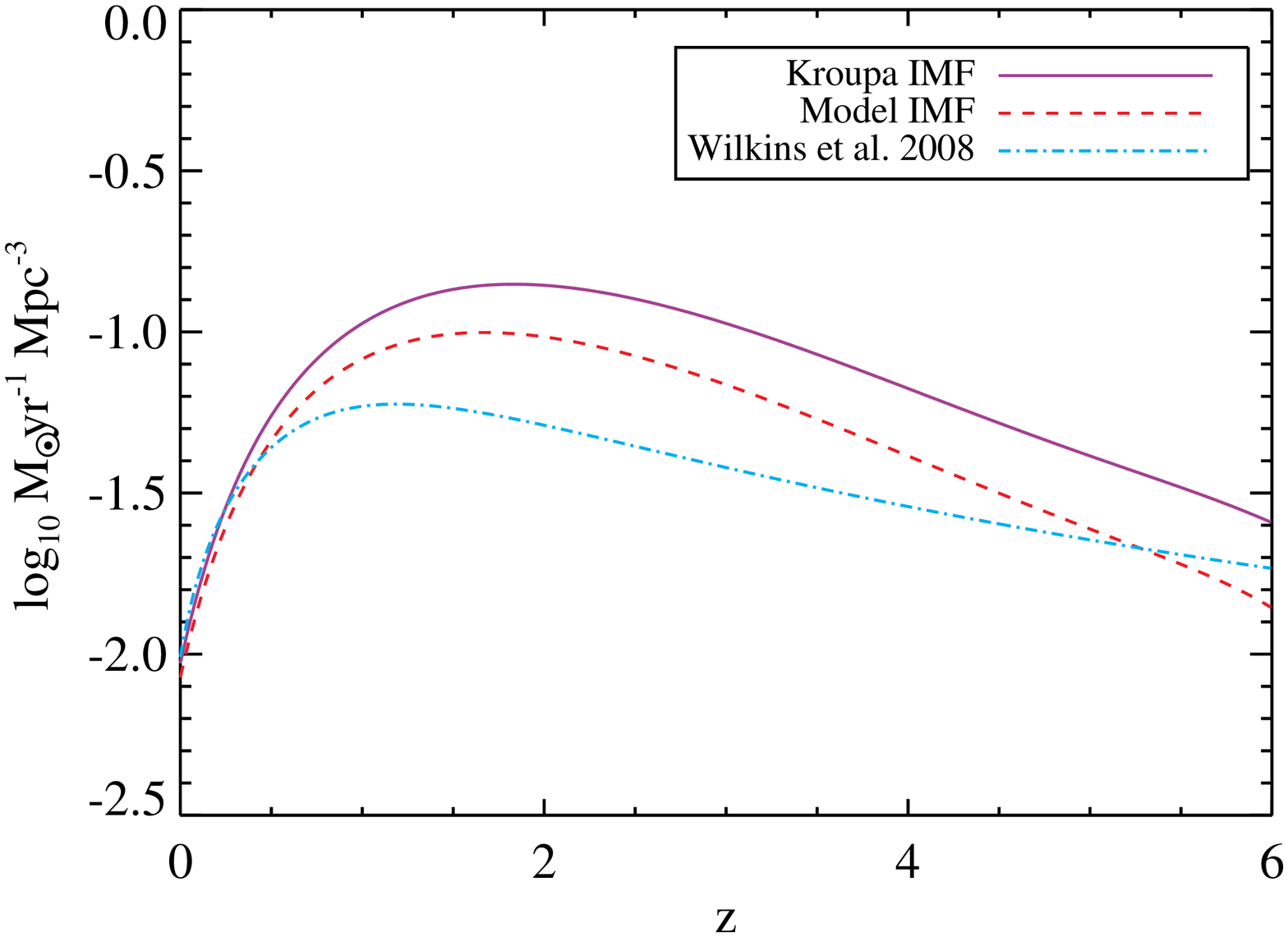}
\caption{Evolution of cosmic star formation rate density.  Purple
  solid line shows model form from \citet{hop10}, assuming a Kroupa
  IMF.  This provides a good fit to the observed SFR density
  measurements of \citep{hop06d}.  When applying our model form for
  IMF variations, the SFR density decreases, as is shown by the red
  dashed line.  The blue dash-dot line shows the \citet{wil08c} SFR
  density that would be necessary to match the observed present-day
  stellar mass density.  The application of our model IMF decreases
  the inferred SFR density.  But because much of the star formation
  occurs in lower-luminosity systems where the IMF is not
  substantially different from canonical, the amount of the decrease
  is not quite enough to account for the discrepancy between the
  integrated SFR density and stellar mass density as presented in
  \citet{wil08c}. \label{figure:madau}}
\end{figure}

Deep surveys have permitted measurements of the evolution of the
star formation rates per unit volume over cosmic time.  Similarly,
increased sensitivities in the near-infrared have allowed for a
more robust reconstruction of the global stellar mass evolution in
galaxies out to \zsim 3.  In principle, the integral of the star
formation rate density, corrected for stellar mass loss, should be
equivalent to the stellar mass density at a given redshift. A number
of authors have noted, however, that there may be a discrepancy
between these two quantities, such that the cosmic SFR density
appears to be overestimated \citep{hop06d,per08,wil08c}.  As alluded
to in \S~\ref{section:introduction}, an IMF weighted toward more
high mass stars at high-\z\ may go some length toward solving this
issue.

Does an IMF that varies with the Jeans mass of GMCs reduce the
inferred SFRs of high-\z\ galaxies enough to remove the discrepancy
between the integrated SFR density and the observed stellar mass
density?  To answer this, we need to determine how a varying IMF would
impact the inferred global SFR density.  To do this, we employ a
functional form for the observed luminosity function of galaxies as a
function of redshift that is quite successful in matching existing
data from \citet{hop10}.  We integrate this functional form as a
function of luminosity to calculate the typical number of galaxies per
Mpc$^3$ as a function of redshift (integrated to 0.1$L^*$).  We
similarly utilise this information to calculate a luminosity and SFR
density as a function of redshift.  This information combined with the
average number density of galaxies as a function of redshift allows us
to calculate the typical SFR per galaxy as a function of redshift.  We
then apply the correction given in Equation~\ref{eq:sfrlir} after
converting to an infrared luminosity via the \citet{ken98a} relations,
and assuming $\lir \approx \lbol$.

Figure~\ref{figure:madau} shows the resulting SFRD evolution assuming
an unevolving Kroupa IMF (purple), versus the model evolving IMF
assuming $M_c\propto\mjchar$ (red dashed).  The unevolving IMF curve
is in good agreement with the SFRD compilation by \citet{hop06d}
when corrected to a Kroupa IMF; for clarity, we do not show the
individual data points.  While the evolving IMF makes a large difference to
the SFRs of very rapidly starforming galaxies, the global effect
is less dramatic, since the SFRD tends to be dominated by more
modestly starforming systems.  Nonetheless, it is clear that it
lowers the SFRD significantly at all $z\ga 1$.

\citet{wil08c} suggested a variable IMF that was empirically
constructed to reconcile the cosmic SFRD evolution with the present-day
stellar mass density.  The resulting SFRD evolution in this model
is shown as the blue dot-dashed line in Figure~\ref{figure:madau}.
The IMF model presented here results in less of a correction to the
present-day stellar mass, since it is clearly closer to the original
unevolving case over much of cosmic time.  Hence it remains to be
seen if the IMF model in this work can actually reconcile these
measurements, although it is clear that it goes some distance towards
this.  The standard Kroupa IMF forms roughly twice as many stars
today as are accounted for in stellar mass density measurements,
whereas our model IMF forms $1.3 \times$ the Wilkins et al. estimate.

This said, an evolving IMF is not necessary to reconcile the
differences between the cosmic SFRD evolution and stellar mass
density.  For example, utilising cosmological SPH simulations,
\citet{cho12} suggest that the two can come into better agreement if
the stellar mass density measurements are missing faint, low-mass
galaxies.  Similarly, \citet{red09} have suggested that the evolution
of the stellar mass density and integral of the cosmic star formation
rate density may not be discrepant. They argue that if one integrates
the luminosity function to a common limit (accounting for the evolution
of $L^*$ with \z), then some of the observed discrepancy between the
star SFRD and stellar mass density can be removed.

Overall, the global SFR density is less affected by our model IMF
than the most rapidly star-forming galaxies, since much of the
global star formation occurs in modestly star-forming systems.  Our
model IMF goes some of the way towards alleviating the putative
discrepancy between the integrated cosmic SFRD and the global stellar
mass evolution, although current observations still disagree on the
magnitude of the discrepancy.

Finally, we note that the rising CMB temperature with redshift sets an
increasing minimum Jeans mass that is non-negligible at large (\zga
3) redshifts.  However, due to the rising SFR per galaxy with
redshift, the average conditions are already warm enough due to
gas-dust coupling that the effect of the CMB is negligible.  As an
example, at \zsim 4, the CMB temperature is $\sim 15$ K.  The mean SFR
per galaxy is $\sim 20 \ \msunyr$, which corresponds to a typical gas
temperature of $\sim 20$ K (see e.g. Figure ~\ref{figure:mjevolution}).


\section{Discussion}
\label{section:discussion}
\subsection{Arguments for and against a varying IMF}

Probably the single greatest piece of evidence for an invariant IMF is
that no conclusive evidence for a varying IMF has ever been seen, even
when probing a fairly wide range of physical conditions and environments
in star forming regions within the Milky Way and nearby galaxies.
For this reason as well as others, all arguments for a varying IMF must
be taken with a healthy degree of skepticism.  The recent review by
\citet{bas10} takes a rather critical view of observational claims of
IMF variations; we refer the reader there for a fuller discussion.

There are many theoretical studies that suggest that the IMF should be
essentially constant in most relevant regimes.  \citet{kru11d} present
a model in which the temperature in typical star forming regions in
GMCs is set by radiation feedback from accretion onto protostars.  In
this scenario, one can arrive at a characteristic stellar mass
primarily via fundamental constants, with only weak scaling with the
physical properties of the GMC.  Similar arguments were made by
\citet{bat09}.  This theory may explain the near constancy of the IMF
in normal star forming regions in the Galaxy, and has the additional
attractive aspect of explaining why stars form to massive enough
scales to begin hydrogen fusion.

On the other hand, there have been a number of recent claims that the
IMF varies with galaxy properties.  It is now reasonably well
established that the H$\alpha$ to UV luminosity ratio varies
systematically with luminosity~\citep{hov08}, SFR~\citep{lee09} and/or
surface brightness~\citep{meu09}, in the sense that for fainter
galaxies there is less H$\alpha$ than expected given the UV
luminosity.  Since H$\alpha$ traces very high mass stars while UV
traces moderately high-mass stars, one possible explanation is that
smaller GMCs are unable to produce very high-mass stars.  In this
scenario, fainter systems with on average smaller GMCs would have an
integrated galactic IMF that is less top-heavy.  This variation would
work in the same direction as the IMF variations considered in this
paper (i.e. more top-heavy/bottom-light in more rapidly star-forming
systems), but for a different reason; also, it is seen to operate down
to very low SFRs (and indeed it is most evident there), whereas our
results suggest that IMF variations based on the Jeans mass conjecture
have little impact at SFRs of the Milky Way or smaller.  That said,
numerous investigations recently have discussed other physical effects
that could lead to the observed trends in H$\alpha$/UV, such as the
leakage of ionizing photons, stochasticity in the formation of the
most massive stars \citep{fum11}, variable star formation histories
\citep{wei12}, and extinction effects.  Similarly, studies
investigating the production rate of ionising photons in stellar
clusters \citep{cal10}, as well as outer galaxy discs \citep{kod12}
find no evidence for a varying IMF.  Hence while intriguing,
H$\alpha$/UV variations are not widely considered to be an unambiguous
indication of a varying IMF.

Early-type galaxies also show tantalizing hints for IMF variations.
\citet{van11} observed the Wing-Ford absorption bands (tracing stellar
masses $\la 0.3M_\odot$) in nearby cluster elliptical galaxies, and
found that they indicate a much larger population of low-mass stars
than expected from a Galactic IMF, i.e. that the IMF in these galaxies
is bottom-heavy.  \citet{cap12} did careful dynamical modeling of
ellipticals from the ATLAS-3D sample, and showed that high-mass
ellipticals tend to have such mass-to-light (M/L) ratios that they can
only be explained by a bottom-heavy IMF.  They also found a strong
trend of IMF variation with M/L.  These data argue for a bottom-heavy
IMF as opposed to a bottom-light one as considered in this paper, but
a naive extrapolation of their trend with M/L suggests that rapidly
star-forming systems may have a top-heavy/bottom-light IMF.  However,
it is also expected that massive star-forming systems at high-$z$
should evolve into present-day passive ellipticals, and thus it is not
immediately clear how to reconcile these differing IMFs within a
single galaxy evolution scenario.  If, indeed, the IMF is bottom-heavy
in the progenitor galaxies of present-day ellipticals, then this
likely rules out the model presented in this paper
\citep{van08,cap12,con12}.

If the IMF is indeed invariant, then what of the cosmic star formation
problems discussed throughout this work?  There are certainly
solutions that do not require an IMF-based solution.  There are many
well-documented systematic uncertainties associated with measuring SFRs
and $M_*$, particularly at high redshifts when only broad-band data is
available~\citep[e.g.][]{sha05}.  It is not difficult to imagine that one
or more of these could alleviate the apparent discrepancies~\citep[see
a discussion in ][]{dav08}.  There could also be failures in current
modeling techniques or assumptions.  For instance, the high sSFR of
$z\sim 2$ galaxies could owe to a strong inefficiency of star formation
in earlier galaxies owing e.g. to lower metallicities~\citep{gne10,bol11},
resulting in an accumulated gas reservoir that fuels excess star formation
at $z\sim 2$~\citep{kru11e}.  This would not alleviate differences between
the observed cosmic SFR and $M_*$ buildup, but perhaps those could be
explained by other systematics; for instance, if the bottom-heavy IMF
seen in early-type galaxies is real, the total cosmic stellar
mass in this component may be underestimated.  Finally, it is worth
noting that numerical uncertainties still abound in modeling galaxy
formation: for instance, \citet{ker11} found that SFRs in ($\sim L^*$)
galaxies modeled with the moving-mesh code \arepo\ are $\sim 2-3$
greater than those modeled using the SPH code \gadget.  With all the
various factor-of-two uncertainties in play, it is difficult to
argue strongly for IMF variations as being the only viable solution to
any given problem.

As discussed in \S~\ref{section:ks}, a varying IMF may bring the
observed Kennicutt-Schmidt relation index down from $\sim 2$ to
$\sim 1.5$, in agreement with most models which suggest a star
formation rate regulated by dynamical processes.  However, as shown
by \citet{ost11}, if the star formation rate is instead regulated
by supernovae-driven turbulence in a medium where the vertical
pressure is dominated by gas, then one would expect a KS index of
$\sim 2$. In this sense, even the reduction of the observed KS index
is not reason enough to require a variable IMF.

\subsection{Arguments for and against the Jeans mass conjecture}

In this work, we make the assumption that the IMF characteristic mass
scales as the GMC Jeans mass.  Here we discuss arguments in favor
of and against this Jeans mass conjecture.

Stars form in dense cores in GMCs \citep[e.g.  ][]{eva99,lad03} that
appear to have a mass function similar in shape to the stellar IMF
\citep{and96,mot01,joh01,nut07}.  This suggests that stars acquire
their mass from the cores they are born in, and are thus plausibly
dependent on the fragmentation scale of the cloud.  In the same vein
denser(less-dense) clouds, which would have a lower(higher) Jeans mass
given the $M_{\rm J} \sim T^{3/2}/n^{1/2}$ scaling, should have a more
bottom-heavy(bottom-light) IMF.  Observations of the Taurus
star-forming region and Orion Nebula Cluster provide some evidence of this effect
\citep{bri02,luh03,luh04,dar12}.

Numerical calculations also suggest that the number of fragments that
form in a GMC is related to the number of Jeans masses available
\citep{bat05}.  \citet{kle07} showed via numerical simulation that
warm gas comparable to what may be found in nuclear starbursts indeed
forms a mass spectrum of collapsed objects with a \mc$\approx
15$~\msunend, though the degree of fragmentation can depend on the
exact form of the equation of state of the gas.  It should be noted
that observations of the Arches \citep{kim06} and Westerlund 1
clusters \citep{bra08,gen11c}, regions similar to those in the
simulations of \citet{kle07} show minimal evidence for a varying IMF.

\citet{elm08} asked, given cooling rates by molecular lines,
and heating rates by dust-gas energy exchange in dense gas, how
should $M_{\rm J}$ scale with density? These authors find a weak
dependence of the Jeans mass on the density as $n^{1/4}$.  Examining
Figure~\ref{figure:mjevolution}, the mass-weighted mean density of
GMCs is $\sim 10^{4} \cmthree$ during the starburst, compared to a few
$\times 10^1 \cmthree$ during quiescent mode.  Following the Elmegreen
et al. scalings\footnote{Note that a similar scaling relation
  comparing to the global SFR is non-trivial to derive.  This depends
  on radiative transfer processes as well as potential contribution to
  the radiation field from old stellar populations.}, this alone would
lead to an increase in Jeans mass of $\sim 4$.  In other words, even with
a relatively weak scaling of $M_{\rm J}$ with density, the conditions
inside nuclear starbursts are extreme enough to result in variations in
the GMC physical properties as found in our models.

A more general argument against the conjecture that $\mc\propto\mjchar$ is
that the choice for the density that sets the Jeans mass is not obvious.
In this work we sidestep this problem by considering only how $M_J$
scales with galaxy properties and assuming that $\mc\propto\mjchar$
without worrying about the constant of proportionality.  However, using
typical numbers from our simulations results in $M_J\sim 250 M_\odot$
for the Milky Way (Figure~\ref{figure:mjvsfr}), which is unsatisfyingly
far from $\mc\approx 0.5 M_\odot$.  \citet{lar05} assumes that the
characteristic density is set by when the gas becomes thermally coupled
to the dust, and argued against this scale being set by turbulent or
magnetic pressure.  But in \citet{kru11d} and other recent models,
the density where isothermality is broken is set instead by radiative
feedback, and thus the IMF characteristic mass is not directly related
to the Jeans mass.  We note that even if the $\mc$ is set by radiative
feedback, the interstellar radiation properties vary substantially among
our various simulated galaxies, and hence there may still be some IMF
variations predicted in such a model.  Nonetheless, the physical origin
of the IMF characteristic mass remains uncertain, and it is by no means
entirely clear that it is directly related to the typical GMC Jeans mass.

In this paper we suggest that heating of \htwo\ gas by young stars
formed during rapid star formation may increase the Jeans mass.  Other
models find similar results, though via different heating mechanisms.
\citet{pap10a} find that the cosmic ray energy densities in starburst
environments may be increased, leading to warmer gas kinetic
temperatures of order $\sim 80-160 $ K.  \citet{pap10b} find that this
can lead to increased Jeans masses up to a factor $\sim 10$.  The
relevant physical processes in the Papadopoulos model are different
from those presented in our work.  In a scenario where cosmic rays
dominate the heating, the gas and dust do not necessarily have to be
thermally coupled.  Nevertheless, the results are quite similar to
what is found in our study.  In \citet{nar11c}, we found that even
when assuming a cosmic ray flux that scales linearly with the star
formation rate, the heating of dust (and subsequent dust-gas energy
exchange at high densities) by young stars in starbursts tends to
dominate the heating of the gas.  Similarly, \citet{hoc10} and
\citet{hoc11} have suggested that a strong X-ray field in the vicinity
of an accreting black hole may impact the fragmentation length of
GMCs, weighting the clump mass spectrum toward higher masses. This
scenario is likely to be most relevant in rather extreme cases
(e.g. near an AGN), and not necessarily representative of quiescently
star-forming galaxies.

Overall, whether the stellar IMF varies in space and/or time, and how
remains an open question.  Strong IMF variations are clearly
disallowed by observations, but mild variations may still be
accommodated within current constraints.  The Jeans mass conjecture is
by no means unassailable, and the resulting mild variations in the IMF
are probably difficult to exclude given current data.

\section{Summary}
\label{section:summary}

We use hydrodynamic galaxy evolution simulations and sophisticated
radiative transfer models to study how the IMF might vary with galaxy
properties, under the assumption that the IMF characteristic mass scales
with the \htwo\ mass-weighted Jeans mass in the ISM.  We find that such an
assumption results in an typical Jeans mass that varies mildly with star
formation rate, roughly as $\mjchar\propto $ \ SFR$^{0.4}$ at star formation
rates above that of the Milky Way.  At SFRs comparable to and lower than
that of the Milky Way, there is little variation in $\mjchar$ in our
model because cosmic ray heating begins to dominate over dust heating,
setting a temperature floor at $\sim 8$~K.  While this variation is not
strong, it is still substantial enough to impact the inferred properties
of rapidly star-forming galaxies.

We parameterise the implied IMF variation by considering its impact on
SFRs derived from bolometric luminosities.  We find that the relation
is well-fit by SFR$=(L_{\rm bol}/10^{10} L_\odot)^{0.88}$, as opposed
to a linear relationship between these quantities as for an invariant
IMF (assuming that any AGN contribution has been excluded).  Since all
extragalactic SFR tracers attempt to measure $L_{\rm bol}$, either
directly in the infrared, via a dust-corrected UV continuum, or via
nebular emission line measures, this would imply that their inferred
SFRs should be lowered in more luminous (i.e. more rapidly
star-forming) systems.  For instance, in sub-millimetre galaxies that
are the most bolometrically luminous systems in the Universe, the
inferred SFRs would be up to a factor of five lower than implied by
canonical conversion factors.  Large, rapidly star-forming disks at
high redshifts would have their inferred SFR lowered by a factor of
two.

We study the impact of such an IMF variation on a variety of galaxy
properties across cosmic time.  These include the star formation law,
the SFR$-M_*$ relation (i.e. the galaxy main sequence), sub-millimetre
galaxies, and the evolution of the cosmic SFR density.  In each case,
previous work has highlighted tantalizing albeit inconclusive evidence
for discrepancies between models and data, or between star formation rate
and stellar mass data.  The discrepancies (if real) all point in the same
direction, in the sense that the SFR inferred from high-mass tracers seems
to exceed the observed growth in stellar mass by $\sim\times 2-3$ during
the peak epoch of cosmic star formation.  In each case, the varying IMF
based on the Jeans mass conjecture tends to go towards reconciling the
possible discrepancies.  We caution that there is no smoking gun evidence
for IMF variations, and there are numerous ways these discrepancies may be
alleviated without varying the IMF.  Nonetheless, it is interesting that
the ansatz of $\mc\propto\mjchar$ makes measurably different predictions
for the evolution of galaxies across cosmic time in a manner that,
at face value, tends to reconcile any possible discrepancies.

We remain agnostic on the validity of our underlying assumption,
namely the Jeans mass conjecture, and we do not assert that IMF
variations are the only solution to the problems discussed here.
Rather, this paper merely explores the ansatz that
$\mc\propto\mjchar$, and then examines its effects on the star
formation rates across cosmic time.  As we discussed in
\S\ref{section:discussion}, current models for the origin of the IMF
characteristic mass are mixed on whether this assumption is correct; a
recent class of successful models argues that \mc \ is set by
radiative feedback in the ISM, not \mjchar.  We leave further
explorations along these lines to a different class of models that
focus on the origin of the IMF.  Additionally, such an evolving IMF
would have a wider range of impacts than explored here, owing to
differences in e.g. stellar mass return, metallicity, and population
synthesis that have not been fully accounted for here.  Also, at
a constant ISM density, the rising CMB temperature will force increasingly
higher Jeans masses with redshift, further complicating
matters. Investigating these effects would require running simulations
that self-consistently model all these processes during their
evolution, which is something we leave for the future.

\section*{Acknowledgements}
The authors thank Mark Krumholz for comments on an early version of
this draft, and Charlie Conroy, Emanuele Daddi, Dusan Keres, Naveen
Reddy and Youngmin Seo for helpful conversations.  This work benefited
from work done and conversations had at the Aspen Center for Physics.
DN acknowledges support from the National Science Foundation via grant
number AST-1009452.  RD was supported by the National Science
Foundation under grant numbers AST-0847667 and AST-0907998.  Computing
resources were obtained through grant number DMS-0619881 from the
National Science Foundation.

\begin{appendix}
\section{A Jeans Mass that Varies with the Median Density}
\label{section:appendix}
\begin{figure}
\hspace{-1cm} \includegraphics[angle=90,scale=0.4]{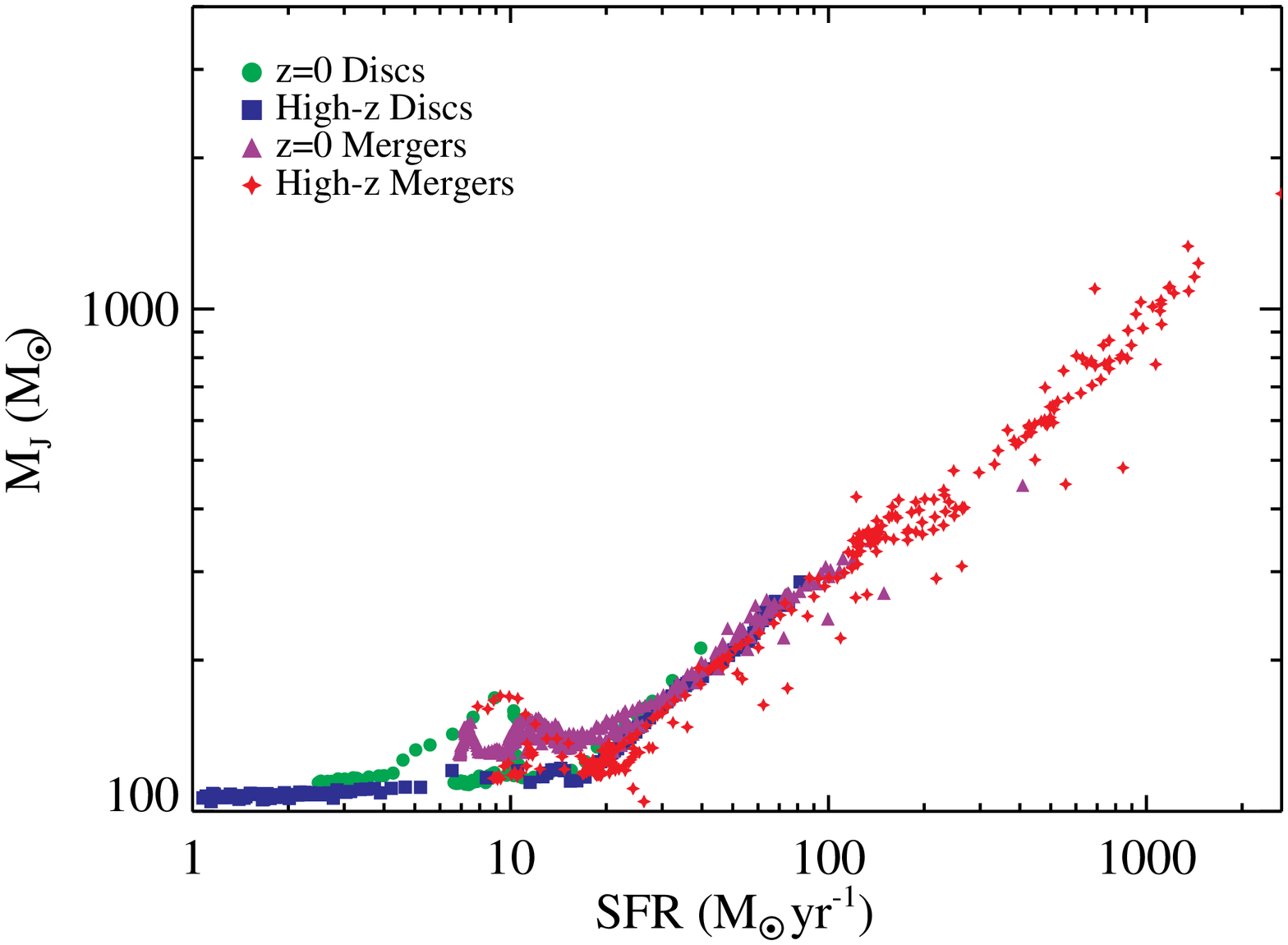}
\caption{Relationship between SFR and Jeans mass when $M_{\rm J}$ is
  calculated using the median density in GMCs, rather than the mean.
  The relation is not dissimilar from using the mean density.  See
  text for details.\label{figure:appendix}}
\end{figure}

It is conceivable that one could consider the density term in the
Jeans mass equation as the median density (i.e.the density above which
half the mass resides), rather than the mean density. As a reminder,
while we resolve the surfaces of GMCs, we do not resolve their
internal structure.  We assume that our GMCs have a lognormal density
distribution.  In this case, the median density can be expressed in
terms of the width of the lognormal, $\sigma_{\rho}$:
\begin{equation}
n_{\rm med} = \bar{n}e^{\sigma_{\rho}^2/2.}
\end{equation}
where $\bar{n}$ is the mean density of the GMC.  Numerical experiments
suggest that $\sigma_{\rho}^2 \approx {\rm ln}(1+3M^2/4)$, where $M$
is the one-dimensional Mach number \citep{ost01,pad02}\footnote{We
  note that other authors have found a range of possible forms for the
  dependence of $\sigma_{\rho}$.  See \citet{lem08} and
  \citet{pri11}.}  In Figure~\ref{figure:appendix}, we show the
relationship between SFR and $M_{\rm J}$ when the Jeans mass is
calculated using $n_{\rm med}$ instead of $\bar{n}$. 

The relation between $M_{\rm J}$ and SFR in
Figure~\ref{figure:appendix} is relatively tight.  When comparing to
Figure~\ref{figure:mjvsfr}, we see that the scatter is greatly reduced
when considering a Jeans mass calculated with the $n_{\rm med}$,
rather than $\bar{n}$.  In Figure~\ref{figure:mjvsfr}, the bulk of the
scatter arises from relatively active systems.  For example, at a SFR
of 50 \msunyrend, the galaxies with the highest average Jeans mass in
Figure~\ref{figure:mjvsfr} are ongoing mergers with warm ISMs, whereas
the galaxies with lower Jeans masses are either mergers during a more
quiescent phase, or discs.  Using the median density rather always
results in a Jeans mass that is lower than when using the mean
density, due to the extra term that scales with $M^2$.  Because
mergers and heavily star-forming discs tend to have high-velocity
dispersion gas \citep[e.g. ][]{nar11b}, their Jeans masses are reduced
substantially when using $n_{\rm med}$.  This tightens the dispersion
dramatically in the range SFR=$10-100$ \msunyrend.

One consequence of using $n_{\rm med}$ rather than $\bar{n}$ is that a
significant increase in the characteristic mass of the IMF is only
seen in heavily star-forming galaxies (SFR $\ga 100 \msunyrend$).  This
suggests that such a variation may only be seen during short-lived
phases in local mergers, or in more heavily star-forming galaxies at
high redshift.  

The results of this paper change little when utilising $n_{\rm med}$
rather than $\bar{n}$ in the Jeans mass calculation.  The
normalisation of the $M_{\rm J}$-SFR relation in
Figure~\ref{figure:appendix} decreases, while the exponent
increases slightly.  These effects cause the normalisation of
Equation~\ref{eq:sfrlir} to decrease by a factor of $\sim 2$, while
the exponent decreases to $\sim 0.85$.

\end{appendix}


\begin{thebibliography}{179}
\expandafter\ifx\csname natexlab\endcsname\relax\def\natexlab#1{#1}\fi

\bibitem[{{Agertz} {et~al.}(2011){Agertz}, {Teyssier}, \& {Moore}}]{age11}
{Agertz}, O., {Teyssier}, R., \& {Moore}, B. 2011, \mnras, 410, 1391

\bibitem[{{Andre} {et~al.}(1996){Andre}, {Ward-Thompson}, \& {Motte}}]{and96}
{Andre}, P., {Ward-Thompson}, D., \& {Motte}, F. 1996, \aap, 314, 625

\bibitem[{{Bastian} {et~al.}(2010){Bastian}, {Covey}, \& {Meyer}}]{bas10}
{Bastian}, N., {Covey}, K.~R., \& {Meyer}, M.~R. 2010, \araa, 48, 339

\bibitem[{{Bate}(2009)}]{bat09}
{Bate}, M.~R. 2009, \mnras, 392, 1363

\bibitem[{{Bate} \& {Bonnell}(2005)}]{bat05}
{Bate}, M.~R. \& {Bonnell}, I.~A. 2005, \mnras, 356, 1201

\bibitem[{{Baugh} {et~al.}(2005)}]{bau05}
{Baugh}, C.~M. {et~al.} 2005, \mnras, 356, 1191

\bibitem[{{Bell} {et~al.}(2007){Bell}, {Zheng}, {Papovich}, {Borch}, {Wolf}, \&
  {Meisenheimer}}]{bel07b}
{Bell}, E.~F., {Zheng}, X.~Z., {Papovich}, C., {Borch}, A., {Wolf}, C., \&
  {Meisenheimer}, K. 2007, \apj, 663, 834

\bibitem[{{Bigiel} {et~al.}(2008){Bigiel}, {Leroy}, {Walter}, {Brinks}, {de
  Blok}, {Madore}, \& {Thornley}}]{big08}
{Bigiel}, F., {Leroy}, A., {Walter}, F., {Brinks}, E., {de Blok}, W.~J.~G.,
  {Madore}, B., \& {Thornley}, M.~D. 2008, \aj, 136, 2846

\bibitem[{{Bolatto} {et~al.}(2011){Bolatto}, {Leroy}, {Jameson}, {Ostriker},
  {Gordon}, {Lawton}, {Stanimirovi{\'c}}, {Israel}, {Madden}, {Hony},
  {Sandstrom}, {Bot}, {Rubio}, {Winkler}, {Roman-Duval}, {van Loon},
  {Oliveira}, \& {Indebetouw}}]{bol11}
{Bolatto}, A.~D., {Leroy}, A.~K., {Jameson}, K., {Ostriker}, E., {Gordon}, K.,
  {Lawton}, B., {Stanimirovi{\'c}}, S., {Israel}, F.~P., {Madden}, S.~C.,
  {Hony}, S., {Sandstrom}, K.~M., {Bot}, C., {Rubio}, M., {Winkler}, P.~F.,
  {Roman-Duval}, J., {van Loon}, J.~T., {Oliveira}, J.~M., \& {Indebetouw}, R.
  2011, \apj, 741, 12

\bibitem[{{Bolatto} {et~al.}(2008){Bolatto}, {Leroy}, {Rosolowsky}, {Walter},
  \& {Blitz}}]{bol08}
{Bolatto}, A.~D., {Leroy}, A.~K., {Rosolowsky}, E., {Walter}, F., \& {Blitz},
  L. 2008, \apj, 686, 948

\bibitem[{{Bothwell} {et~al.}(2010)}]{bot09}
{Bothwell}, M.~S. {et~al.} 2010, \mnras, 405, 219

\bibitem[{{Bouch{\'e}} {et~al.}(2007)}]{bou07}
{Bouch{\'e}}, N. {et~al.} 2007, \apj, 671, 303

\bibitem[{{Bournaud} {et~al.}(2010){Bournaud}, {Elmegreen}, {Teyssier},
  {Block}, \& {Puerari}}]{bou10}
{Bournaud}, F., {Elmegreen}, B.~G., {Teyssier}, R., {Block}, D.~L., \&
  {Puerari}, I. 2010, \mnras, 409, 1088

\bibitem[{{Bouwens} {et~al.}(2011){Bouwens}, {Illingworth}, {Oesch}, {Franx},
  {Labbe}, {Trenti}, {van Dokkum}, {Carollo}, {Gonzalez}, \& {Magee}}]{bou11b}
{Bouwens}, R.~J., {Illingworth}, G.~D., {Oesch}, P.~A., {Franx}, M., {Labbe},
  I., {Trenti}, M., {van Dokkum}, P., {Carollo}, C.~M., {Gonzalez}, V., \&
  {Magee}, D. 2011, arXiv/1109.0994

\bibitem[{{Brandner} {et~al.}(2008){Brandner}, {Clark}, {Stolte}, {Waters},
  {Negueruela}, \& {Goodwin}}]{bra08}
{Brandner}, W., {Clark}, J.~S., {Stolte}, A., {Waters}, R., {Negueruela}, I.,
  \& {Goodwin}, S.~P. 2008, \aap, 478, 137

\bibitem[{{Brice{\~n}o} {et~al.}(2002){Brice{\~n}o}, {Luhman}, {Hartmann},
  {Stauffer}, \& {Kirkpatrick}}]{bri02}
{Brice{\~n}o}, C., {Luhman}, K.~L., {Hartmann}, L., {Stauffer}, J.~R., \&
  {Kirkpatrick}, J.~D. 2002, \apj, 580, 317

\bibitem[{{Bullock} {et~al.}(2001){Bullock}, {Kolatt}, {Sigad}, {Somerville},
  {Kravtsov}, {Klypin}, {Primack}, \& {Dekel}}]{bul01}
{Bullock}, J.~S., {Kolatt}, T.~S., {Sigad}, Y., {Somerville}, R.~S.,
  {Kravtsov}, A.~V., {Klypin}, A.~A., {Primack}, J.~R., \& {Dekel}, A. 2001,
  \mnras, 321, 559

\bibitem[{{Bussmann} {et~al.}(2008){Bussmann}, {Narayanan}, {Shirley},
  {Juneau}, {Wu}, {Solomon}, {Vanden Bout}, {Moustakas}, \& {Walker}}]{bus08}
{Bussmann}, R.~S., {Narayanan}, D., {Shirley}, Y.~L., {Juneau}, S., {Wu}, J.,
  {Solomon}, P.~M., {Vanden Bout}, P.~A., {Moustakas}, J., \& {Walker}, C.~K.
  2008, \apjl, 681, L73

\bibitem[{{Calura} {et~al.}(2008){Calura}, {Pipino}, \& {Matteucci}}]{cal08}
{Calura}, F., {Pipino}, A., \& {Matteucci}, F. 2008, \aap, 479, 669

\bibitem[{{Calzetti} {et~al.}(2010){Calzetti}, {Chandar}, {Lee}, {Elmegreen},
  {Kennicutt}, \& {Whitmore}}]{cal10}
{Calzetti}, D., {Chandar}, R., {Lee}, J.~C., {Elmegreen}, B.~G., {Kennicutt},
  R.~C., \& {Whitmore}, B. 2010, \apjl, 719, L158

\bibitem[{{Cappellari} {et~al.}(2012){Cappellari}, {McDermid}, {Alatalo},
  {Blitz}, {Bois}, {Bournaud}, {Bureau}, {Crocker}, {Davies}, {Davis}, {de
  Zeeuw}, {Duc}, {Emsellem}, {Khochfar}, {Krajnovic}, {Kuntschner},
  {Lablanche}, {Morganti}, {Naab}, {Oosterloo}, {Sarzi}, {Scott}, {Serra},
  {Weijmans}, \& {Young}}]{cap12}
{Cappellari}, M., {McDermid}, R.~M., {Alatalo}, K., {Blitz}, L., {Bois}, M.,
  {Bournaud}, F., {Bureau}, M., {Crocker}, A.~F., {Davies}, R.~L., {Davis},
  T.~A., {de Zeeuw}, P.~T., {Duc}, P.-A., {Emsellem}, E., {Khochfar}, S.,
  {Krajnovic}, D., {Kuntschner}, H., {Lablanche}, P.-Y., {Morganti}, R.,
  {Naab}, T., {Oosterloo}, T., {Sarzi}, M., {Scott}, N., {Serra}, P.,
  {Weijmans}, A.-M., \& {Young}, L.~M. 2012, arXiv/1202.3308

\bibitem[{{Ceverino} {et~al.}(2010{\natexlab{a}}){Ceverino}, {Dekel}, \&
  {Bournaud}}]{cer10}
{Ceverino}, D., {Dekel}, A., \& {Bournaud}, F. 2010{\natexlab{a}}, \mnras, 440

\bibitem[{{Ceverino} {et~al.}(2010{\natexlab{b}}){Ceverino}, {Dekel}, \&
  {Bournaud}}]{cev10}
---. 2010{\natexlab{b}}, \mnras, 404, 2151

\bibitem[{{Choi} \& {Nagamine}(2012)}]{cho12}
{Choi}, J.-H. \& {Nagamine}, K. 2012, \mnras, 419, 1280

\bibitem[{{Conroy} \& {Gunn}(2010)}]{con10b}
{Conroy}, C. \& {Gunn}, J.~E. 2010, \apj, 712, 833

\bibitem[{{Conroy} {et~al.}(2009){Conroy}, {Gunn}, \& {White}}]{con09b}
{Conroy}, C., {Gunn}, J.~E., \& {White}, M. 2009, \apj, 699, 486

\bibitem[{{Conroy} \& {van Dokkum}(2012)}]{con12}
{Conroy}, C. \& {van Dokkum}, P. 2012, \apj, 747, 69

\bibitem[{{Conroy} {et~al.}(2010){Conroy}, {White}, \& {Gunn}}]{con10a}
{Conroy}, C., {White}, M., \& {Gunn}, J.~E. 2010, \apj, 708, 58

\bibitem[{{Cox} {et~al.}(2006)}]{cox06b}
{Cox}, T.~J. {et~al.} 2006, \apj, 650, 791

\bibitem[{{Crosthwaite} \& {Turner}(2007)}]{cro07}
{Crosthwaite}, L.~P. \& {Turner}, J.~L. 2007, \aj, 134, 1827

\bibitem[{{Da Rio} {et~al.}(2012){Da Rio}, {Robberto}, {Hillenbrand},
  {Henning}, \& {Stassun}}]{dar12}
{Da Rio}, N., {Robberto}, M., {Hillenbrand}, L.~A., {Henning}, T., \&
  {Stassun}, K.~G. 2012, \apj, 748, 14

\bibitem[{{Daddi} {et~al.}(2005)}]{dad05}
{Daddi}, E. {et~al.} 2005, \apjl, 631, L13

\bibitem[{{Daddi} {et~al.}(2007)}]{dad07}
---. 2007, \apj, 670, 156

\bibitem[{{Daddi} {et~al.}(2010{\natexlab{a}})}]{dad10b}
---. 2010{\natexlab{a}}, \apjl, 714, L118

\bibitem[{{Daddi} {et~al.}(2010{\natexlab{b}})}]{dad10a}
---. 2010{\natexlab{b}}, \apj, 713, 686

\bibitem[{{Dav{\'e}}(2008)}]{dav08}
{Dav{\'e}}, R. 2008, \mnras, 385, 147

\bibitem[{{Dav{\'e}}(2011)}]{dav11c}
{Dav{\'e}}, R. 2011, in Astronomical Society of the Pacific Conference Series,
  Vol. 440, UP2010: Have Observations Revealed a Variable Upper End of the
  Initial Mass Function?, ed. {M.~Treyer, T.~Wyder, J.~Neill, M.~Seibert, \&
  J.~Lee}, 353

\bibitem[{{Dav{\'e}} {et~al.}(2010){Dav{\'e}}, {Finlator}, {Oppenheimer},
  {Fardal}, {Katz}, {Kere{\v s}}, \& {Weinberg}}]{dav10}
{Dav{\'e}}, R., {Finlator}, K., {Oppenheimer}, B.~D., {Fardal}, M., {Katz}, N.,
  {Kere{\v s}}, D., \& {Weinberg}, D.~H. 2010, \mnras, 404, 1355

\bibitem[{{Dav{\'e}} {et~al.}(2000){Dav{\'e}}, {Gardner}, {Hernquist}, {Katz},
  \& {Weinberg}}]{dav00}
{Dav{\'e}}, R., {Gardner}, J., {Hernquist}, L., {Katz}, N., \& {Weinberg}, D.
  2000, in Astronomical Society of the Pacific Conference Series, Vol. 200,
  Clustering at High Redshift, ed. {A.~Mazure, O.~Le F{\`e}vre, \& V.~Le Brun},
  173

\bibitem[{{Dav{\'e}} {et~al.}(2011){Dav{\'e}}, {Oppenheimer}, \&
  {Finlator}}]{dav11b}
{Dav{\'e}}, R., {Oppenheimer}, B.~D., \& {Finlator}, K. 2011, \mnras, 415, 11

\bibitem[{{Dekel} {et~al.}(2009{\natexlab{a}}){Dekel}, {Birnboim}, {Engel},
  {Freundlich}, {Goerdt}, {Mumcuoglu}, {Neistein}, {Pichon}, {Teyssier}, \&
  {Zinger}}]{dek08}
{Dekel}, A., {Birnboim}, Y., {Engel}, G., {Freundlich}, J., {Goerdt}, T.,
  {Mumcuoglu}, M., {Neistein}, E., {Pichon}, C., {Teyssier}, R., \& {Zinger},
  E. 2009{\natexlab{a}}, \nat, 457, 451

\bibitem[{{Dekel} {et~al.}(2009{\natexlab{b}}){Dekel}, {Birnboim}, {Engel},
  {Freundlich}, {Goerdt}, {Mumcuoglu}, {Neistein}, {Pichon}, {Teyssier}, \&
  {Zinger}}]{dek09}
---. 2009{\natexlab{b}}, \nat, 457, 451

\bibitem[{{Draine} \& {Li}(2007)}]{dra07}
{Draine}, B.~T. \& {Li}, A. 2007, \apj, 657, 810

\bibitem[{{Dwek}(1998)}]{dwe98}
{Dwek}, E. 1998, \apj, 501, 643

\bibitem[{{Elbaz} {et~al.}(2007){Elbaz}, {Daddi}, {Le Borgne}, {Dickinson},
  {Alexander}, {Chary}, {Starck}, {Brandt}, {Kitzbichler}, {MacDonald},
  {Nonino}, {Popesso}, {Stern}, \& {Vanzella}}]{elb07}
{Elbaz}, D., {Daddi}, E., {Le Borgne}, D., {Dickinson}, M., {Alexander}, D.~M.,
  {Chary}, R.-R., {Starck}, J.-L., {Brandt}, W.~N., {Kitzbichler}, M.,
  {MacDonald}, E., {Nonino}, M., {Popesso}, P., {Stern}, D., \& {Vanzella}, E.
  2007, \aap, 468, 33

\bibitem[{{Elmegreen}(2002)}]{elm02a}
{Elmegreen}, B.~G. 2002, \apj, 577, 206

\bibitem[{{Elmegreen} {et~al.}(2008){Elmegreen}, {Klessen}, \&
  {Wilson}}]{elm08}
{Elmegreen}, B.~G., {Klessen}, R.~S., \& {Wilson}, C.~D. 2008, \apj, 681, 365

\bibitem[{{Elsner} {et~al.}(2008){Elsner}, {Feulner}, \& {Hopp}}]{els08}
{Elsner}, F., {Feulner}, G., \& {Hopp}, U. 2008, \aap, 477, 503

\bibitem[{{Engel} {et~al.}(2010){Engel}, {Tacconi}, {Davies}, {Neri}, {Smail},
  {Chapman}, {Genzel}, {Cox}, {Greve}, {Ivison}, {Blain}, {Bertoldi}, \&
  {Omont}}]{eng10}
{Engel}, H., {Tacconi}, L.~J., {Davies}, R.~I., {Neri}, R., {Smail}, I.,
  {Chapman}, S.~C., {Genzel}, R., {Cox}, P., {Greve}, T.~R., {Ivison}, R.~J.,
  {Blain}, A., {Bertoldi}, F., \& {Omont}, A. 2010, \apj, 724, 233

\bibitem[{{Evans}(1999)}]{eva99}
{Evans}, II, N.~J. 1999, \araa, 37, 311

\bibitem[{{Evans} {et~al.}(2009){Evans}, {Dunham}, {J{\o}rgensen}, {Enoch},
  {Mer{\'{\i}}n}, {van Dishoeck}, {Alcal{\'a}}, {Myers}, {Stapelfeldt},
  {Huard}, {Allen}, {Harvey}, {van Kempen}, {Blake}, {Koerner}, {Mundy},
  {Padgett}, \& {Sargent}}]{eva09}
{Evans}, II, N.~J., {Dunham}, M.~M., {J{\o}rgensen}, J.~K., {Enoch}, M.~L.,
  {Mer{\'{\i}}n}, B., {van Dishoeck}, E.~F., {Alcal{\'a}}, J.~M., {Myers},
  P.~C., {Stapelfeldt}, K.~R., {Huard}, T.~L., {Allen}, L.~E., {Harvey}, P.~M.,
  {van Kempen}, T., {Blake}, G.~A., {Koerner}, D.~W., {Mundy}, L.~G.,
  {Padgett}, D.~L., \& {Sargent}, A.~I. 2009, \apjs, 181, 321

\bibitem[{{Fakhouri} \& {Ma}(2008)}]{fak08}
{Fakhouri}, O. \& {Ma}, C.-P. 2008, \mnras, 386, 577

\bibitem[{{Fardal} {et~al.}(2007){Fardal}, {Katz}, {Weinberg}, \&
  {Dav{\'e}}}]{far07}
{Fardal}, M.~A., {Katz}, N., {Weinberg}, D.~H., \& {Dav{\'e}}, R. 2007, \mnras,
  379, 985

\bibitem[{{Feldmann} {et~al.}(2011){Feldmann}, {Gnedin}, \& {Kravtsov}}]{fel11}
{Feldmann}, R., {Gnedin}, N.~Y., \& {Kravtsov}, A.~V. 2011, \apj, 732, 115

\bibitem[{{Finlator} {et~al.}(2006){Finlator}, {Dav{\'e}}, {Papovich}, \&
  {Hernquist}}]{fin06}
{Finlator}, K., {Dav{\'e}}, R., {Papovich}, C., \& {Hernquist}, L. 2006, \apj,
  639, 672

\bibitem[{{F{\"o}rster Schreiber} {et~al.}(2003){F{\"o}rster Schreiber},
  {Genzel}, {Lutz}, \& {Sternberg}}]{for03}
{F{\"o}rster Schreiber}, N.~M., {Genzel}, R., {Lutz}, D., \& {Sternberg}, A.
  2003, \apj, 599, 193

\bibitem[{{Fukui} \& {Kawamura}(2010)}]{fuk10}
{Fukui}, Y. \& {Kawamura}, A. 2010, \araa, 48, 547

\bibitem[{{Fumagalli} {et~al.}(2011){Fumagalli}, {da Silva}, \&
  {Krumholz}}]{fum11}
{Fumagalli}, M., {da Silva}, R.~L., \& {Krumholz}, M.~R. 2011, \apjl, 741, L26

\bibitem[{{Gao} \& {Solomon}(2004{\natexlab{a}})}]{gao04a}
{Gao}, Y. \& {Solomon}, P.~M. 2004{\natexlab{a}}, \apjs, 152, 63

\bibitem[{{Gao} \& {Solomon}(2004{\natexlab{b}})}]{gao04b}
---. 2004{\natexlab{b}}, \apj, 606, 271

\bibitem[{{Garcia-Burillo} {et~al.}(2011){Garcia-Burillo}, {Usero},
  {Alonso-Herrero}, {Gracia-Carpio}, {Pereira-Santaella}, {Colina}, {Planesas},
  \& {Arribas}}]{gar11}
{Garcia-Burillo}, S., {Usero}, A., {Alonso-Herrero}, A., {Gracia-Carpio}, J.,
  {Pereira-Santaella}, M., {Colina}, L., {Planesas}, P., \& {Arribas}, S. 2011,
  arXiv/1111.6773

\bibitem[{{Gennaro} {et~al.}(2011){Gennaro}, {Brandner}, {Stolte}, \&
  {Henning}}]{gen11c}
{Gennaro}, M., {Brandner}, W., {Stolte}, A., \& {Henning}, T. 2011, \mnras,
  412, 2469

\bibitem[{{Genzel} {et~al.}(2010)}]{gen10}
{Genzel}, R. {et~al.} 2010, \mnras, 407, 2091

\bibitem[{{Genzel} {et~al.}(2011)}]{gen11b}
---. 2011, arXiv/1106.2098

\bibitem[{{Gnedin} \& {Kravtsov}(2010)}]{gne10}
{Gnedin}, N.~Y. \& {Kravtsov}, A.~V. 2010, \apj, 714, 287

\bibitem[{{Gonz{\'a}lez} {et~al.}(2011){Gonz{\'a}lez}, {Lacey}, {Baugh}, \&
  {Frenk}}]{gon11}
{Gonz{\'a}lez}, J.~E., {Lacey}, C.~G., {Baugh}, C.~M., \& {Frenk}, C.~S. 2011,
  \mnras, 413, 749

\bibitem[{{Gonz{\'a}lez} {et~al.}(2010){Gonz{\'a}lez}, {Labb{\'e}}, {Bouwens},
  {Illingworth}, {Franx}, {Kriek}, \& {Brammer}}]{gon10}
{Gonz{\'a}lez}, V., {Labb{\'e}}, I., {Bouwens}, R.~J., {Illingworth}, G.,
  {Franx}, M., {Kriek}, M., \& {Brammer}, G.~B. 2010, \apj, 713, 115

\bibitem[{{Governato} {et~al.}(2009){Governato}, {Brook}, {Brooks}, {Mayer},
  {Willman}, {Jonsson}, {Stilp}, {Pope}, {Christensen}, {Wadsley}, \&
  {Quinn}}]{gov09}
{Governato}, F., {Brook}, C.~B., {Brooks}, A.~M., {Mayer}, L., {Willman}, B.,
  {Jonsson}, P., {Stilp}, A.~M., {Pope}, L., {Christensen}, C., {Wadsley}, J.,
  \& {Quinn}, T. 2009, \mnras, 398, 312

\bibitem[{{Greve} {et~al.}(2005)}]{gre05}
{Greve}, T.~R. {et~al.} 2005, \mnras, 359, 1165

\bibitem[{{Guo} \& {White}(2008)}]{guo08}
{Guo}, Q. \& {White}, S.~D.~M. 2008, \mnras, 384, 2

\bibitem[{{Hayward} {et~al.}(2011){Hayward}, {Kere{\v s}}, {Jonsson},
  {Narayanan}, {Cox}, \& {Hernquist}}]{hay11}
{Hayward}, C.~C., {Kere{\v s}}, D., {Jonsson}, P., {Narayanan}, D., {Cox},
  T.~J., \& {Hernquist}, L. 2011, arXiv/1101.0002

\bibitem[{{Hayward} {et~al.}(2010){Hayward}, {Narayanan}, {Jonsson}, {Cox},
  {Kere{\v s}}, {Hopkins}, \& {Hernquist}}]{hay10}
{Hayward}, C.~C., {Narayanan}, D., {Jonsson}, P., {Cox}, T.~J., {Kere{\v s}},
  D., {Hopkins}, P.~F., \& {Hernquist}, L. 2010, Conference Proceedings for
  UP2010: Have Observations Revealed a Variable Upper End of the Initial Mass
  Function? Treyer, Lee, Seibert, Wyder, Neil eds. arXiv/1008.4584

\bibitem[{{Heiderman} {et~al.}(2010){Heiderman}, {Evans}, {Allen}, {Huard}, \&
  {Heyer}}]{hei10}
{Heiderman}, A., {Evans}, II, N.~J., {Allen}, L.~E., {Huard}, T., \& {Heyer},
  M. 2010, \apj, 723, 1019

\bibitem[{{Hernquist}(1990)}]{her90}
{Hernquist}, L. 1990, \apj, 356, 359

\bibitem[{{Hocuk} \& {Spaans}(2010)}]{hoc10}
{Hocuk}, S. \& {Spaans}, M. 2010, \aap, 522, A24

\bibitem[{{Hocuk} \& {Spaans}(2011)}]{hoc11}
---. 2011, \aap, 536, A41

\bibitem[{{Hopkins} \& {Beacom}(2006)}]{hop06d}
{Hopkins}, A.~M. \& {Beacom}, J.~F. 2006, \apj, 651, 142

\bibitem[{{Hopkins} {et~al.}(2010){Hopkins}, {Younger}, {Hayward}, {Narayanan},
  \& {Hernquist}}]{hop10}
{Hopkins}, P.~F., {Younger}, J.~D., {Hayward}, C.~C., {Narayanan}, D., \&
  {Hernquist}, L. 2010, \mnras, 402, 1693

\bibitem[{{Hoversten} \& {Glazebrook}(2008)}]{hov08}
{Hoversten}, E.~A. \& {Glazebrook}, K. 2008, \apj, 675, 163

\bibitem[{{Iono} {et~al.}(2009){Iono}, {Wilson}, {Yun}, {Baker}, {Petitpas},
  {Peck}, {Krips}, {Cox}, {Matsushita}, {Mihos}, \& {Pihlstrom}}]{ion09}
{Iono}, D., {Wilson}, C.~D., {Yun}, M.~S., {Baker}, A.~J., {Petitpas}, G.~R.,
  {Peck}, A.~B., {Krips}, M., {Cox}, T.~J., {Matsushita}, S., {Mihos}, J.~C.,
  \& {Pihlstrom}, Y. 2009, \apj, 695, 1537

\bibitem[{{Johnstone} {et~al.}(2001){Johnstone}, {Fich}, {Mitchell}, \&
  {Moriarty-Schieven}}]{joh01}
{Johnstone}, D., {Fich}, M., {Mitchell}, G.~F., \& {Moriarty-Schieven}, G.
  2001, \apj, 559, 307

\bibitem[{{Jonsson}(2006)}]{jon06a}
{Jonsson}, P. 2006, \mnras, 372, 2

\bibitem[{{Jonsson} {et~al.}(2010){Jonsson}, {Groves}, \& {Cox}}]{jon10a}
{Jonsson}, P., {Groves}, B.~A., \& {Cox}, T.~J. 2010, \mnras, 186

\bibitem[{{Jonsson} \& {Primack}(2010)}]{jon10b}
{Jonsson}, P. \& {Primack}, J.~R. 2010, New Astronomy, 15, 509

\bibitem[{{Kang} {et~al.}(2010){Kang}, {Lin}, {Skibba}, \& {Chen}}]{kan10}
{Kang}, X., {Lin}, W.~P., {Skibba}, R., \& {Chen}, D.~N. 2010, \apj, 713, 1301

\bibitem[{{Kennicutt}(1998{\natexlab{a}})}]{ken98a}
{Kennicutt}, Jr., R.~C. 1998{\natexlab{a}}, \araa, 36, 189

\bibitem[{{Kennicutt}(1998{\natexlab{b}})}]{ken98b}
---. 1998{\natexlab{b}}, \apj, 498, 541

\bibitem[{{Kennicutt} {et~al.}(2007){Kennicutt}, {Calzetti}, {Walter}, {Helou},
  {Hollenbach}, {Armus}, {Bendo}, {Dale}, {Draine}, {Engelbracht}, \&
  {Gordon}}]{ken07}
{Kennicutt}, Jr., R.~C., {Calzetti}, D., {Walter}, F., {Helou}, G.,
  {Hollenbach}, D.~J., {Armus}, L., {Bendo}, G., {Dale}, D.~A., {Draine},
  B.~T., {Engelbracht}, C.~W., \& {Gordon}, K.~D. 2007, \apj, 671, 333

\bibitem[{{Kennicutt} \& {Evans}(2012)}]{ken12}
{Kennicutt}, Jr., R.~C. \& {Evans}, II, N.~J. 2012, arXiv/1204.3552

\bibitem[{{Keres} {et~al.}(2011){Keres}, {Vogelsberger}, {Sijacki}, {Springel},
  \& {Hernquist}}]{ker11}
{Keres}, D., {Vogelsberger}, M., {Sijacki}, D., {Springel}, V., \& {Hernquist},
  L. 2011, arXiv/1109.4638

\bibitem[{{Kere{\v s}} {et~al.}(2009){Kere{\v s}}, {Katz}, {Fardal},
  {Dav{\'e}}, \& {Weinberg}}]{ker09}
{Kere{\v s}}, D., {Katz}, N., {Fardal}, M., {Dav{\'e}}, R., \& {Weinberg},
  D.~H. 2009, \mnras, 395, 160

\bibitem[{{Kere{\v s}} {et~al.}(2005){Kere{\v s}}, {Katz}, {Weinberg}, \&
  {Dav{\'e}}}]{ker05}
{Kere{\v s}}, D., {Katz}, N., {Weinberg}, D.~H., \& {Dav{\'e}}, R. 2005,
  \mnras, 363, 2

\bibitem[{{Kim} {et~al.}(2006){Kim}, {Figer}, {Kudritzki}, \&
  {Najarro}}]{kim06}
{Kim}, S.~S., {Figer}, D.~F., {Kudritzki}, R.~P., \& {Najarro}, F. 2006, \apjl,
  653, L113

\bibitem[{{Klessen} {et~al.}(2007){Klessen}, {Spaans}, \& {Jappsen}}]{kle07}
{Klessen}, R.~S., {Spaans}, M., \& {Jappsen}, A.-K. 2007, \mnras, 374, L29

\bibitem[{{Koda} {et~al.}(2012){Koda}, {Yagi}, {Boissier}, {Gil de Paz},
  {Imanishi}, {Donovan Meyer}, {Madore}, \& {Thilker}}]{kod12}
{Koda}, J., {Yagi}, M., {Boissier}, S., {Gil de Paz}, A., {Imanishi}, M.,
  {Donovan Meyer}, J., {Madore}, B.~F., \& {Thilker}, D.~A. 2012, \apj, 749, 20

\bibitem[{{Kroupa}(2002)}]{kro02}
{Kroupa}, P. 2002, Science, 295, 82

\bibitem[{{Krumholz}(2011)}]{kru11d}
{Krumholz}, M.~R. 2011, \apj, 743, 110

\bibitem[{{Krumholz} \& {Dekel}(2011)}]{kru11e}
{Krumholz}, M.~R. \& {Dekel}, A. 2011, arXiv/1106.0301

\bibitem[{{Krumholz} {et~al.}(2011{\natexlab{a}}){Krumholz}, {Dekel}, \&
  {McKee}}]{kru11c}
{Krumholz}, M.~R., {Dekel}, A., \& {McKee}, C.~F. 2011{\natexlab{a}},
  arXiv/1109.4150

\bibitem[{{Krumholz} {et~al.}(2012){Krumholz}, {Dekel}, \& {McKee}}]{kru12a}
---. 2012, \apj, 745, 69

\bibitem[{{Krumholz} {et~al.}(2011{\natexlab{b}}){Krumholz}, {Leroy}, \&
  {McKee}}]{kru11a}
{Krumholz}, M.~R., {Leroy}, A.~K., \& {McKee}, C.~F. 2011{\natexlab{b}}, \apj,
  731, 25

\bibitem[{{Krumholz} \& {McKee}(2005)}]{kru05}
{Krumholz}, M.~R. \& {McKee}, C.~F. 2005, \apj, 630, 250

\bibitem[{{Krumholz} {et~al.}(2008){Krumholz}, {McKee}, \& {Tumlinson}}]{kru08}
{Krumholz}, M.~R., {McKee}, C.~F., \& {Tumlinson}, J. 2008, \apj, 689, 865

\bibitem[{{Krumholz} {et~al.}(2009{\natexlab{a}}){Krumholz}, {McKee}, \&
  {Tumlinson}}]{kru09a}
---. 2009{\natexlab{a}}, \apj, 693, 216

\bibitem[{{Krumholz} {et~al.}(2009{\natexlab{b}}){Krumholz}, {McKee}, \&
  {Tumlinson}}]{kru09b}
---. 2009{\natexlab{b}}, \apj, 699, 850

\bibitem[{{Krumholz} \& {Tan}(2007)}]{kru07b}
{Krumholz}, M.~R. \& {Tan}, J.~C. 2007, \apj, 654, 304

\bibitem[{{Krumholz} \& {Thompson}(2007)}]{kru07}
{Krumholz}, M.~R. \& {Thompson}, T.~A. 2007, \apj, 669, 289

\bibitem[{{Lada} {et~al.}(2011){Lada}, {Forbrich}, {Lombardi}, \&
  {Alves}}]{lad11}
{Lada}, C.~J., {Forbrich}, J., {Lombardi}, M., \& {Alves}, J.~F. 2011,
  arXiv/1112.4466

\bibitem[{{Lada} \& {Lada}(2003)}]{lad03}
{Lada}, C.~J. \& {Lada}, E.~A. 2003, \araa, 41, 57

\bibitem[{{Larson}(2005)}]{lar05}
{Larson}, R.~B. 2005, \mnras, 359, 211

\bibitem[{{Lee} {et~al.}(2009)}]{lee09}
{Lee}, J.~C. {et~al.} 2009, \apj, 706, 599

\bibitem[{{Leitherer} {et~al.}(1999)}]{lei99}
{Leitherer}, C. {et~al.} 1999, \apjs, 123, 3

\bibitem[{{Lemaster} \& {Stone}(2008)}]{lem08}
{Lemaster}, M.~N. \& {Stone}, J.~M. 2008, \apjl, 682, L97

\bibitem[{{Leroy} {et~al.}(2011)}]{ler11}
{Leroy}, A.~K. {et~al.} 2011, arXiv/1102.4618

\bibitem[{{Luhman}(2004)}]{luh04}
{Luhman}, K.~L. 2004, \apj, 617, 1216

\bibitem[{{Luhman} {et~al.}(2003){Luhman}, {Brice{\~n}o}, {Stauffer},
  {Hartmann}, {Barrado y Navascu{\'e}s}, \& {Caldwell}}]{luh03}
{Luhman}, K.~L., {Brice{\~n}o}, C., {Stauffer}, J.~R., {Hartmann}, L., {Barrado
  y Navascu{\'e}s}, D., \& {Caldwell}, N. 2003, \apj, 590, 348

\bibitem[{{Madau} {et~al.}(1996){Madau}, {Ferguson}, {Dickinson}, {Giavalisco},
  {Steidel}, \& {Fruchter}}]{mad96}
{Madau}, P., {Ferguson}, H.~C., {Dickinson}, M.~E., {Giavalisco}, M.,
  {Steidel}, C.~C., \& {Fruchter}, A. 1996, \mnras, 283, 1388

\bibitem[{{Magnelli} {et~al.}(2012)}]{mag12}
{Magnelli}, B. {et~al.} 2012, arXiv/1202.0761

\bibitem[{{McKee} \& {Ostriker}(2007)}]{mck07}
{McKee}, C.~F. \& {Ostriker}, E.~C. 2007, \araa, 45, 565

\bibitem[{{McKee} \& {Ostriker}(1977)}]{mck77}
{McKee}, C.~F. \& {Ostriker}, J.~P. 1977, \apj, 218, 148

\bibitem[{{McLure} {et~al.}(2011){McLure}, {Dunlop}, {de Ravel}, {Cirasuolo},
  {Ellis}, {Schenker}, {Robertson}, {Koekemoer}, {Stark}, \& {Bowler}}]{mcl11}
{McLure}, R.~J., {Dunlop}, J.~S., {de Ravel}, L., {Cirasuolo}, M., {Ellis},
  R.~S., {Schenker}, M., {Robertson}, B.~E., {Koekemoer}, A.~M., {Stark},
  D.~P., \& {Bowler}, R.~A.~A. 2011, \mnras, 418, 2074

\bibitem[{{Meurer} {et~al.}(2009)}]{meu09}
{Meurer}, G.~R. {et~al.} 2009, \apj, 695, 765

\bibitem[{{Mo} {et~al.}(1998){Mo}, {Mao}, \& {White}}]{mo98}
{Mo}, H.~J., {Mao}, S., \& {White}, S.~D.~M. 1998, \mnras, 295, 319

\bibitem[{{Motte} {et~al.}(2001){Motte}, {Andr{\'e}}, {Ward-Thompson}, \&
  {Bontemps}}]{mot01}
{Motte}, F., {Andr{\'e}}, P., {Ward-Thompson}, D., \& {Bontemps}, S. 2001,
  \aap, 372, L41

\bibitem[{{Narayanan} {et~al.}(2011{\natexlab{a}}){Narayanan}, {Cox},
  {Hayward}, \& {Hernquist}}]{nar11a}
{Narayanan}, D., {Cox}, T.~J., {Hayward}, C.~C., \& {Hernquist}, L.
  2011{\natexlab{a}}, \mnras, 412, 287

\bibitem[{{Narayanan} {et~al.}(2005){Narayanan}, {Groppi}, {Kulesa}, \&
  {Walker}}]{nar05}
{Narayanan}, D., {Groppi}, C.~E., {Kulesa}, C.~A., \& {Walker}, C.~K. 2005,
  \apj, 630, 269

\bibitem[{{Narayanan} {et~al.}(2010){Narayanan}, {Hayward}, {Cox}, {Hernquist},
  {Jonsson}, {Younger}, \& {Groves}}]{nar10a}
{Narayanan}, D., {Hayward}, C.~C., {Cox}, T.~J., {Hernquist}, L., {Jonsson},
  P., {Younger}, J.~D., \& {Groves}, B. 2010, \mnras, 401, 1613

\bibitem[{{Narayanan} {et~al.}(2011{\natexlab{b}}){Narayanan}, {Krumholz},
  {Ostriker}, \& {Hernquist}}]{nar11b}
{Narayanan}, D., {Krumholz}, M., {Ostriker}, E.~C., \& {Hernquist}, L.
  2011{\natexlab{b}}, \mnras, 418, 664

\bibitem[{{Narayanan} {et~al.}(2011{\natexlab{c}}){Narayanan}, {Krumholz},
  {Ostriker}, \& {Hernquist}}]{nar11c}
{Narayanan}, D., {Krumholz}, M.~R., {Ostriker}, E.~C., \& {Hernquist}, L.
  2011{\natexlab{c}}, arXiv/1110.3791

\bibitem[{{Nayakshin} \& {Sunyaev}(2005)}]{nay05}
{Nayakshin}, S. \& {Sunyaev}, R. 2005, \mnras, 364, L23

\bibitem[{{Niemi} {et~al.}(2012){Niemi}, {Somerville}, {Ferguson}, {Huang},
  {Lotz}, \& {Koekemoer}}]{nie12}
{Niemi}, S.-M., {Somerville}, R.~S., {Ferguson}, H.~C., {Huang}, K.-H., {Lotz},
  J., \& {Koekemoer}, A.~M. 2012, arXiv/1201.2410

\bibitem[{{Noeske} {et~al.}(2007{\natexlab{a}})}]{noe07b}
{Noeske}, K.~G. {et~al.} 2007{\natexlab{a}}, \apjl, 660, L47

\bibitem[{{Noeske} {et~al.}(2007{\natexlab{b}})}]{noe07a}
---. 2007{\natexlab{b}}, \apjl, 660, L43

\bibitem[{{Nutter} \& {Ward-Thompson}(2007)}]{nut07}
{Nutter}, D. \& {Ward-Thompson}, D. 2007, \mnras, 374, 1413

\bibitem[{{Ostriker} \& {Shetty}(2011)}]{ost11}
{Ostriker}, E.~C. \& {Shetty}, R. 2011, \apj, 731, 41

\bibitem[{{Ostriker} {et~al.}(2001){Ostriker}, {Stone}, \& {Gammie}}]{ost01}
{Ostriker}, E.~C., {Stone}, J.~M., \& {Gammie}, C.~F. 2001, \apj, 546, 980

\bibitem[{{Padoan} \& {Nordlund}(2002)}]{pad02}
{Padoan}, P. \& {Nordlund}, {\AA}. 2002, \apj, 576, 870

\bibitem[{{Papadopoulos}(2010)}]{pap10a}
{Papadopoulos}, P.~P. 2010, \apj, 720, 226

\bibitem[{{Papadopoulos} {et~al.}(2011){Papadopoulos}, {Thi}, {Miniati}, \&
  {Viti}}]{pap10b}
{Papadopoulos}, P.~P., {Thi}, W.-F., {Miniati}, F., \& {Viti}, S. 2011, \mnras,
  414, 1705

\bibitem[{{P{\'e}rez-Gonz{\'a}lez} {et~al.}(2008){P{\'e}rez-Gonz{\'a}lez},
  {Rieke}, {Villar}, {Barro}, {Blaylock}, {Egami}, {Gallego}, {Gil de Paz},
  {Pascual}, {Zamorano}, \& {Donley}}]{per08}
{P{\'e}rez-Gonz{\'a}lez}, P.~G., {Rieke}, G.~H., {Villar}, V., {Barro}, G.,
  {Blaylock}, M., {Egami}, E., {Gallego}, J., {Gil de Paz}, A., {Pascual}, S.,
  {Zamorano}, J., \& {Donley}, J.~L. 2008, \apj, 675, 234

\bibitem[{{Price} {et~al.}(2011){Price}, {Federrath}, \& {Brunt}}]{pri11}
{Price}, D.~J., {Federrath}, C., \& {Brunt}, C.~M. 2011, \apjl, 727, L21+

\bibitem[{{Reddy} \& {Steidel}(2009)}]{red09}
{Reddy}, N.~A. \& {Steidel}, C.~C. 2009, \apj, 692, 778

\bibitem[{{Rieke} {et~al.}(1993){Rieke}, {Loken}, {Rieke}, \&
  {Tamblyn}}]{rie93}
{Rieke}, G.~H., {Loken}, K., {Rieke}, M.~J., \& {Tamblyn}, P. 1993, \apj, 412,
  99

\bibitem[{{Robertson} {et~al.}(2004){Robertson}, {Yoshida}, {Springel}, \&
  {Hernquist}}]{rob04}
{Robertson}, B., {Yoshida}, N., {Springel}, V., \& {Hernquist}, L. 2004, \apj,
  606, 32

\bibitem[{{Robertson} {et~al.}(2006)}]{rob06a}
{Robertson}, B. {et~al.} 2006, \apj, 641, 21

\bibitem[{{Robertson} \& {Kravtsov}(2008)}]{rob08}
{Robertson}, B.~E. \& {Kravtsov}, A.~V. 2008, \apj, 680, 1083

\bibitem[{{Rodighiero} {et~al.}(2011)}]{rod11}
{Rodighiero}, G. {et~al.} 2011, \apjl, 739, L40

\bibitem[{{Schruba} {et~al.}(2011)}]{sch11}
{Schruba}, A. {et~al.} 2011, \aj, 142, 37

\bibitem[{{Schuster} {et~al.}(2007){Schuster}, {Kramer}, {Hitschfeld},
  {Garcia-Burillo}, \& {Mookerjea}}]{sch07}
{Schuster}, K.~F., {Kramer}, C., {Hitschfeld}, M., {Garcia-Burillo}, S., \&
  {Mookerjea}, B. 2007, \aap, 461, 143

\bibitem[{{Shapley}(2011)}]{sha11}
{Shapley}, A.~E. 2011, ARA\&A in press; arXiv/1107.5060

\bibitem[{{Shapley} {et~al.}(2005){Shapley}, {Steidel}, {Erb}, {Reddy},
  {Adelberger}, {Pettini}, {Barmby}, \& {Huang}}]{sha05}
{Shapley}, A.~E., {Steidel}, C.~C., {Erb}, D.~K., {Reddy}, N.~A., {Adelberger},
  K.~L., {Pettini}, M., {Barmby}, P., \& {Huang}, J. 2005, \apj, 626, 698

\bibitem[{{Shetty} \& {Ostriker}(2008)}]{she08}
{Shetty}, R. \& {Ostriker}, E.~C. 2008, \apj, 684, 978

\bibitem[{{Shetty} {et~al.}(2011{\natexlab{a}})}]{she11b}
{Shetty}, R. {et~al.} 2011{\natexlab{a}}, arXiv/1104.3695

\bibitem[{{Shetty} {et~al.}(2011{\natexlab{b}})}]{she11a}
---. 2011{\natexlab{b}}, \mnras, 412, 1686

\bibitem[{{Silk}(1997)}]{sil97}
{Silk}, J. 1997, \apj, 481, 703

\bibitem[{{Sobral} {et~al.}(2012){Sobral}, {Smail}, {Best}, {Geach}, {Matsuda},
  {Stott}, {Cirasuolo}, \& {Kurk}}]{sob12}
{Sobral}, D., {Smail}, I., {Best}, P.~N., {Geach}, J.~E., {Matsuda}, Y.,
  {Stott}, J.~P., {Cirasuolo}, M., \& {Kurk}, J. 2012, arXiv/1202.3436

\bibitem[{{Springel}(2000)}]{spr00}
{Springel}, V. 2000, \mnras, 312, 859

\bibitem[{{Springel}(2005)}]{spr05b}
---. 2005, \mnras, 364, 1105

\bibitem[{{Springel} {et~al.}(2005){Springel}, {Di Matteo}, \&
  {Hernquist}}]{spr05a}
{Springel}, V., {Di Matteo}, T., \& {Hernquist}, L. 2005, \mnras, 361, 776

\bibitem[{{Springel} \& {Hernquist}(2003)}]{spr03a}
{Springel}, V. \& {Hernquist}, L. 2003, \mnras, 339, 289

\bibitem[{{Stark} {et~al.}(2009){Stark}, {Ellis}, {Bunker}, {Bundy}, {Targett},
  {Benson}, \& {Lacy}}]{sta09}
{Stark}, D.~P., {Ellis}, R.~S., {Bunker}, A., {Bundy}, K., {Targett}, T.,
  {Benson}, A., \& {Lacy}, M. 2009, \apj, 697, 1493

\bibitem[{{Stolte} {et~al.}(2005){Stolte}, {Brandner}, {Grebel}, {Lenzen}, \&
  {Lagrange}}]{sto05}
{Stolte}, A., {Brandner}, W., {Grebel}, E.~K., {Lenzen}, R., \& {Lagrange},
  A.-M. 2005, \apjl, 628, L113

\bibitem[{{Tacconi} {et~al.}(2006)}]{tac06}
{Tacconi}, L.~J. {et~al.} 2006, \apj, 640, 228

\bibitem[{{Tacconi} {et~al.}(2008)}]{tac08}
---. 2008, \apj, 680, 246

\bibitem[{{Tan}(2010)}]{tan10}
{Tan}, J.~C. 2010, \apjl, 710, L88

\bibitem[{{Tumlinson}(2007)}]{tum07}
{Tumlinson}, J. 2007, \apjl, 664, L63

\bibitem[{{van Dokkum}(2008)}]{van08}
{van Dokkum}, P.~G. 2008, \apj, 674, 29

\bibitem[{{van Dokkum} \& {Conroy}(2011)}]{van11}
{van Dokkum}, P.~G. \& {Conroy}, C. 2011, \apjl, 735, L13

\bibitem[{{V{\'a}zquez} \& {Leitherer}(2005)}]{vaz05}
{V{\'a}zquez}, G.~A. \& {Leitherer}, C. 2005, \apj, 621, 695

\bibitem[{{Vladilo}(1998)}]{vla98}
{Vladilo}, G. 1998, \apj, 493, 583

\bibitem[{{Watson}(2011)}]{wat11}
{Watson}, D. 2011, arXiv/1107.6031

\bibitem[{{Weingartner} \& {Draine}(2001)}]{wei01}
{Weingartner}, J.~C. \& {Draine}, B.~T. 2001, \apj, 548, 296

\bibitem[{{Weisz} {et~al.}(2012){Weisz}, {Johnson}, {Johnson}, {Skillman},
  {Lee}, {Kennicutt}, {Calzetti}, {van Zee}, {Bothwell}, {Dalcanton}, {Dale},
  \& {Williams}}]{wei12}
{Weisz}, D.~R., {Johnson}, B.~D., {Johnson}, L.~C., {Skillman}, E.~D., {Lee},
  J.~C., {Kennicutt}, R.~C., {Calzetti}, D., {van Zee}, L., {Bothwell}, M.~S.,
  {Dalcanton}, J.~J., {Dale}, D.~A., \& {Williams}, B.~F. 2012, \apj, 744, 44

\bibitem[{{Wilkins} {et~al.}(2008){Wilkins}, {Trentham}, \& {Hopkins}}]{wil08c}
{Wilkins}, S.~M., {Trentham}, N., \& {Hopkins}, A.~M. 2008, \mnras, 385, 687

\bibitem[{{Wolfire} {et~al.}(2010){Wolfire}, {Hollenbach}, \& {McKee}}]{wol10}
{Wolfire}, M.~G., {Hollenbach}, D., \& {McKee}, C.~F. 2010, \apj, 716, 1191

\bibitem[{{Wong} \& {Blitz}(2002)}]{won02}
{Wong}, T. \& {Blitz}, L. 2002, \apj, 569, 157

\bibitem[{{Wu} {et~al.}(2010){Wu}, {Evans}, {Shirley}, \& {Knez}}]{wu10}
{Wu}, J., {Evans}, N., {Shirley}, Y., \& {Knez}, C. 2010, arXiv/1004.0398

\bibitem[{{Wu} {et~al.}(2005){Wu}, {Evans}, {Gao}, {Solomon}, {Shirley}, \&
  {Vanden Bout}}]{wu05}
{Wu}, J., {Evans}, II, N.~J., {Gao}, Y., {Solomon}, P.~M., {Shirley}, Y.~L., \&
  {Vanden Bout}, P.~A. 2005, \apjl, 635, L173

\bibitem[{{Wuyts} {et~al.}(2011)}]{wuy11}
{Wuyts}, S. {et~al.} 2011, \apj, 742, 96

\end{thebibliography}


\newpage
\end{document}